\documentclass[twocolumn,a4paper,reprint,amssymb,aps,prd,groupedaddress,superscriptaddress]{revtex4-1}
\pdfoutput=1
\usepackage{dcolumn}
\usepackage{array}
\usepackage{xcolor}
\usepackage{amsmath}
\usepackage{graphicx,bm,color,psfrag}
\usepackage{subfigure,bm}
\usepackage{changes}
\usepackage{amsfonts}
\usepackage{lipsum}
\usepackage{enumitem}  
\usepackage{booktabs}
\usepackage{array}
\usepackage{hyperref} 
\usepackage{changes}

\begin{document}
\title{Rastall gravity extension of the standard $\Lambda$CDM model:\\ theoretical features and observational constraints}

\author{\"{O}zg\"{u}r Akarsu}
\email{akarsuo@itu.edu.tr}
\affiliation{Department of Physics, Istanbul Technical University, Maslak 34469 Istanbul, Turkey}

\author{Nihan Kat{\i}rc{\i}}
\email{nihan.katirci@itu.edu.tr}
\affiliation{Department of Physics, Istanbul Technical University, Maslak 34469 Istanbul, Turkey}

\author{Suresh Kumar}
\email{suresh.kumar@pilani.bits-pilani.ac.in}
\affiliation{Department of Mathematics, BITS Pilani, Pilani Campus, Rajasthan-333031, India}

\author{Rafael C. Nunes}
\email{rafadcnunes@gmail.com}
\affiliation{Divis\~ao de Astrof\'isica, Instituto Nacional de Pesquisas Espaciais, Avenida dos Astronautas 1758, S\~ao Jos\'e dos Campos, 12227-010, SP, Brazil}

\author{Burcu \"{O}zt\"{u}rk}
\email{ozturkburc@itu.edu.tr}
\affiliation{Department of Physics, Istanbul Technical University, Maslak 34469 Istanbul, Turkey}

\author{Shivani Sharma}
\email{p20170025@pilani.bits-pilani.ac.in}
\affiliation{Department of Mathematics, BITS Pilani, Pilani Campus, Rajasthan-333031, India}

\begin{abstract}
We present a detailed investigation of the Rastall gravity extension of the standard $\Lambda$CDM model. We review the model for two simultaneous modifications of different nature in the Friedmann equation due to the Rastall gravity: the new contributions of the material (actual) sources (considered as effective source) and the altered evolution of the material sources. We discuss the role/behavior of these modifications with regard to some low redshift tensions, including the so-called $H_0$ tension, prevailing within the standard $\Lambda$CDM. We constrain the model at the level of linear perturbations, and obtain the first constraints through a robust and accurate analysis using the latest full Planck cosmic microwave background (CMB) data, with and without including baryon acoustic oscillations (BAO) data. We find that the Rastall parameter $\epsilon$ (null for general relativity) is consistent with zero at 68\% CL (with a tendency towards positive values, $-0.0001 < \epsilon < 0.0007$ (CMB+BAO) at 68\% CL), which in turn implies no significant statistical evidence for deviation from general relativity, and also a precision of $\mathcal{O}(10^{-4})$ for the coefficient $-1/2$ of the term $g_{\mu\nu}R$ in the Einstein field equations of general relativity (guaranteeing the local energy-momentum conservation). We explore the consequences led by the Rastall gravity on the cosmological parameters in the light of the observational analyses. It turns out that the effective source, with a present-day density parameter $\Omega_{\rm X0}=-0.0010\pm0.0013$ (CMB+BAO, 68\% CL), dynamically screens the usual vacuum energy at high redshifts, but this mechanism barely works due to the opposition by the altered evolution of cold dark matter. Consequently, two simultaneous modifications of different nature in the Friedmann equation by the Rastall gravity act against each other, and do not help to considerably relax the low redshift tensions, including the so-called $H_0$ tension. Our results may offer a guide for the research community that studies the Rastall gravity in various aspects of gravitation and cosmology.

\end{abstract}
\maketitle

\section{Introduction}
\label{sec:intro}
The most successful description of the dynamics and the large-scale structure of the Universe via excellent agreement with a wide variety of the currently available data \cite{Riess:1998cb,Ade:2015xua,Alam:2016hwk,Abbott:2017wau,Aghanim:2018eyx}, in the literature so far, is known to be presented by
the standard (base) Lambda cold dark matter ($\Lambda$CDM) model that relies on the inflationary paradigm \cite{Starobinsky:1980te,Guth:1980zm,Linde:1981mu,Albrecht:1982wi}. Despite its great success, in addition to the notoriously challenging theoretical issues related to $\Lambda$ (cosmological constant) \cite{Weinberg:1988cp,Sahni:1999gb,Peebles:2002gy,Padmanabhan:2002ji}, it has recently begun to suffer from some persistent tensions of various degrees of significance between some existing data sets (see, e.g., \cite{tension02,tension03,Zhao:2017cud,Bullock:2017xww,Freedman:2017yms} for further reading). Such tensions have received immense attention as these could be signaling new physics beyond the well established fundamental theories of physics underpinning the standard $\Lambda$CDM model. For example, the value of the Hubble constant $H_{0}$ predicted by the cosmic microwave background (CMB) Planck data \cite{Ade:2015xua,Aghanim:2018eyx} within the framework of the standard $\Lambda$CDM model is in serious disagreement with the direct local distance ladder measurements \cite{Riess:2016jrr,Riess:2018byc,Riess:2019cxk,Freedman:2019jwv}. This tension becomes even more compelling as it worsens (relieves only partially) when the $\Lambda$ is replaced by the simplest minimally coupled single-field quintessence (phantom or quintom models), see \cite{Vagnozzi:2018jhn,Colgain:2019joh,DiValentino:2019exe,DiValentino:2019dzu} and \cite{Visinelli:2019qqu} for further references. Surprisingly, it has been reported that the $H_0$ tension---as well as a number of other persistent low-redshift tensions---may be alleviated by a dynamical dark energy (DE) whose density can assume negative values or vanish at high redshifts \cite{Delubac:2014aqe,Aubourg:2014yra,Sahni:2014ooa,Mortsell:2018mfj,Poulin:2018zxs,Capozziello:2018jya,Wang:2018fng,Dutta:2018vmq,Banihashemi:2018oxo,Banihashemi:2018has,Visinelli:2019qqu,Akarsu:2019hmw,Ye:2020btb,Perez:2020cwa}. The fact that the CMB Planck data favor positive spatial curvature---which imitates a negative energy density source---on top of the standard $\Lambda$CDM model in contrast to the inflationary paradigm might also be signaling a need for such DE sources \cite{Aghanim:2018eyx} (see also \cite{Kumar:2015,DiValentino:2019qzk,DiValentino:2020hov}).

The constraint on $H_0$ from CMB data is inferred by analyzing the separation/distances in the acoustic peak positions, which depend on the angular scale $\theta_{\star} = r_{\star}/D_{\rm M}$, where $r_{\star}=\int^{\infty}_{z_{\star}}c_sH^{-1}{\rm d}z$ is the comoving sound horizon at recombination---the comoving distance traveled by a sound wave from the reheating to the recombination epoch determined by the pre-recombination ($z>z_{\star}$) physics---and $D_{\rm M}=\int_0^{z_{\star}}H^{-1}{\rm d}z$ is the comoving angular diameter distance at recombination---determined by the post-recombination physics, viz., by the $H(z)$ for $z<z_{\star}$ \cite{Dodelson03}. As DE (generically, described by the equation of state (EoS) of the form $p\sim-\rho$) is negligible at large redshifts, it does not affect the pre-recombination physics, viz., $r_{\star}$. Also, Planck satellite measures $\theta_{\star}$ very robustly and almost independently of the cosmological model with a high precision \cite{Aghanim:2018eyx}. Therefore, $D_{\rm M}$ must remain the same in different DE models (with an exception of the Early Dark Energy models \cite{Poulin:2018cxd}, which modify $r_{\star}$). Thus, a DE assuming negative energy density values at large redshifts---when it is still not negligible (say when $z\lesssim 2$)---will lead to a decrease in $H(z)$ compared to the standard $\Lambda$CDM model at large redshifts (for $z\gtrsim 2$), more than the phantom/quintom DE models could achieve. The compensation of this decrease, viz., keeping $D_{\rm M}$ unaltered, then implies an enhanced $H(z)$ for $z\lesssim2$, i.e., $H_0$ as well. Moreover, the sign change in the DE density can even lead to a non-monotonic behavior of $H(z)$, say, at $z\sim 2$, which in turn may reconcile such models with the Ly-$\alpha $ forest measurement of the baryon acoustic oscillations (BAO) data as well. Indeed, the DE that assumes negative density values at large redshift came to the agenda first when it turned out that, within the standard $\Lambda$CDM model, the Ly-$\alpha $ forest measurement by the BOSS collaboration prefers a smaller value of the dust density parameter compared to the value preferred by the CMB data \cite{Delubac:2014aqe}. They then reported a clear detection of DE consistent with $\Lambda>0$ for $z<1$, but with a preference for negative energy density values for $z>1.6$, and argued that this Ly-$\alpha $ data from $z\approx 2.34$ can be described by a non-monotonic evolution of $H(z)$, which is difficult to achieve in any model with non-negative DE density \cite{Aubourg:2014yra}. The Planck collaboration excludes the Ly-$\alpha$ data from their default BAO compilation as it persistently remains in large tension with the standard $\Lambda$CDM model \cite{Aghanim:2018eyx}. They argue, in line with \cite{Aubourg:2014yra}, that it is difficult to construct well-motivated extensions to the standard $\Lambda$CDM model that can resolve the tension with the Ly-$\alpha$ data, and suggest further work to assess whether this is a statistical fluctuation caused by small systematic errors, or is a signature of new physics. Of course, an \textit{actual} (\textit{physical}) DE source with a negative density would be physically ill, which might be pointing a necessity of considering modified theories of gravity (see \cite{Copeland:2006wr,Caldwell:2009ix,Clifton:2011jh,DeFelice:2010aj,Capozziello:2011et,Nojiri:2017ncd,Nojiri:2010wj} for reviews on DE and modified theories of gravity), from which an \textit{effective} DE source with desired features could be defined. In line with all these, it was argued in \cite{Sahni:2014ooa} that the Ly-$\alpha$ data could be in tension not only with the standard $\Lambda$CDM model but also with standard DE models---restricted to positive energy density values and based on the standard Friedmann equation of general relativity (GR)---, which might be implying (i) DE assuming negative energy density values at high redshifts, (ii) there is a non-conservation of matter source, or (iii) the standard Friedmann equation of GR is inadequate since one could be dealing with a modified gravity theory. An example of (iii) is provided by models in which the $\Lambda$ is screened (or compensated) by a dynamically evolving counter-term, which arises in the Friedmann equation due to the modified gravity. Such examples are in fact familiar from an effective source defined by the collection of all modifications to the usual Einstein field equations in scalar-tensor theories of gravitation, namely, when the cosmological gravitational coupling strength gets weaker with increasing redshift \cite{Boisseau:2000pr,Sahni:2006pa}. See, e.g., Refs. \cite{Umilta:2015cta,Ballardini:2016cvy,Akarsu:2019pvi,Rossi:2019lgt} suggesting larger $H_0$ values in the light of observational analyses when the Brans-Dicke theory or its extensions are considered. There is a  wide range of examples, related to these three possibilities, that exist (i) in theories in which $\Lambda $ relaxes from a large initial value via an adjustment mechanism \cite{Dolgov:1982qq,Bauer:2010wj}, (ii) in the models in which $\Lambda$ itself spontaneously switches sign \cite{Franchino-Vinas:2019nqy,Akarsu:2019hmw}, (iii) in cosmological models based on Gauss-Bonnet gravity \cite{Zhou:2009cy}, (iv) in braneworld models \cite{Sahni:2002dx,Brax:2003fv}, (v) in loop quantum cosmology \cite{Ashtekar:2006wn,Ashtekar:2011ni}, (vi) in higher-dimensional cosmologies that accommodate dynamical reduction of the internal space \cite{Chodos:1979vk,Dereli:1982ar,Akarsu:2012vv,Russo:2018akp}, (vii) in generalisations of the form of the matter Lagrangian in a non-linear way \cite{Akarsu:2017ohj,Board:2017ign,Akarsu:2019ygx}, (viii) in some constructions within the unimodular gravity violating the local energy-momentum conservation law \cite{Perez:2020cwa}.

In this paper, we carry out a detailed theoretical and observational investigation of the Rastall gravity \cite{Rastall:1972,Rastall:1976uh}, which presents a simple mathematical generalization of GR leading to physically rich features that could be related to the cosmological points discussed above. As it was pointed out in \cite{Visser:2017gpz}, an equivalent of a cosmological model constructed within the Rastall gravity can always be achieved by introducing a particular non-minimal coupling between the cosmological constant and matter stresses within the usual GR, which, however, could be neither trivial nor economical (see Sect.~\ref{sec:GRequiv}). Although the Rastall gravity presents a simple generalization of GR (derived from Einstein-Hilbert action) at the level of the field equations, there is no consensus on that it could be derived from an action of a well established fundamental theory, but some attempts in this direction have been made. It was shown in \cite{Smalley84} that its field equations can be derived from a variational principle in a Weyl-Cartan theory, in which the metricity condition for the connection is dropped and the torsion is allowed. Some Lagrangian formulations for it have been proposed within the framework of $f(R,T)$ and $f(R,\mathcal{L}_m)$ gravities \cite{DeMoraes:2019mef,Shabani:2020wja}.

The original physical idea behind the Rastall gravity was the fact that the tight constraints on the local energy-momentum conservation law ($\nabla^{\mu}T_{\mu\nu}=0$) in flat-spacetime do not necessarily imply that this law is valid in curved-spacetime (e.g., in an expanding Universe on cosmological scales) as well \cite{Rastall:1972,Rastall:1976uh}. This idea was then followed by the introduction of the relation $\nabla^{\mu}T_{\mu\nu}\propto \nabla_{\nu}R$ between the energy-momentum tensor (EMT) and the scalar curvature (the simplest curvature invariant of a Riemannian manifold). This leads to a simple mathematical generalization of the standard Einstein field equations of GR adding the term $g_{\mu\nu}R$ to the field equations with an arbitrary coefficient, the so-called Rastall gravity: $R_{\mu\nu}-\alpha g_{\mu\nu}R= T_{\mu\nu}$ with $\alpha$ being a real constant. The particular case $\alpha=\frac{1}{2}$ leading to the Einstein tensor $G_{\mu\nu}$ of GR is unique as it, through the twice-contracted Bianchi identity, yields $\nabla^{\mu}G_{\mu\nu}=0$, and therefore guarantees the local conservation of the EMT of the total material content, i.e., $\nabla^{\mu}T_{\mu\nu}=0$. Therefore, any deviation from $\alpha=\frac{1}{2}$ (GR) will lead to two simultaneous modifications of different nature in the standard Einstein field equations: (i) The new term $\epsilon g_{\mu\nu}R$ (with $\epsilon=\alpha-\frac{1}{2}$) of the form of the usual vacuum energy of quantum field theory (QFT) ($T_{\mu\nu}=g_{\mu\nu}\Lambda$) appears in the spacetime geometry side of the Einstein field equations of GR. (ii) The evolution of the energy density of an actual material source gets altered from its usual one in GR, in a certain way owing to the non-conservation of the EMT described by $\nabla^{\mu}T_{\mu\nu}=-\epsilon\nabla_{\nu}R$. These two simultaneous features tempted us to carefully study the extension of the standard $\Lambda$CDM model replacing the gravity theory from GR to Rastall gravity due to the following reasons: The new term $\epsilon g_{\mu\nu}R$ could dynamically screen the usual vacuum energy at high redshifts for a certain range of $\epsilon$ as suggested, e.g., in \cite{Sahni:2014ooa}. In addition, this extension could also modify inverse proportionality of the dust (e.g., the CDM) energy density to the comoving volume scale factor. It is not clear whether these two simultaneous modifications in the Friedmann equation will support or act against each other, and then whether these together could address the low redshift tensions.

 The Rastall gravity has been attracting a lot of attention by the communities in the field of gravitation and cosmology in the recent years. See, for instance, \cite{Heydarzade:2017wxu} for black hole solutions, \cite{Ziaie:2019jfl} for gravitational collapse, \cite{Bamba:2018zil} for thermodynamic analysis, and \cite{Batista:2010nq,Batista:2011,Moradpour17,Daouda:2018kuo} for some cosmological applications. It has been suggested in \cite{Moradpour17} that, when the Renyi entropy of non-extensive systems is attributed to the horizon of spatially flat Friedmann-Robertson-Walker Universe in Rastall gravity, the late time acceleration can be generated from the non-conservation of dust. It was argued in \cite{Daouda:2018kuo} that Rastall theory provides a proper platform for generalizing the unimodular gravity (the trace-free Einstein gravity) \cite{Unruh:1988in,Ellis:2010uc}, wherein the usual vacuum energy does not gravitate but the cosmological constant arises as an integration constant. The authors of Ref. \cite{Batista:2011} impose the Rastall gravity contributions from the non-conservation of dust upon DE which makes it clustering. Then, it is suggested in \cite{Batista:2012hv,Fabris15} that this model resembles the standard $\Lambda$CDM model at the background level (with an $\epsilon$ not strictly constrained, provided that the DE yields an equation of state parameter very close to minus unity), while in \cite{Khyllep:2019odd} that the evolution of the growth index displays a significant deviation from that in the standard $\Lambda$CDM model. One of the main motivations behind such attempts is that observational signatures of non-conservation in the dark sector is expected in the non-linear regime on intermediate or small scales, and is not inconsistent with the currently available cosmological data (see \cite{vandeBruck:2019vzd} for details). In fact, just upon the proposal of the Rastall gravity, the Rastall parameter has been quite tightly constrained from local physics, relying basically on the non-conservation property of the model, to be $|\epsilon|\lesssim 10^{-15}$, which suggests that the Rastall gravity deviates from GR only negligibly and thereby it is rather unattractive \cite{Lindblom:1982}. Using realistic equations of state for the neutron star interior, an astrophysical constraint is placed on the Rastall gravity suggests that it is well consistent with GR as $|\epsilon|\lesssim10^{-2}$ \cite{Oliveira:2015lka}. In contrast, the study \cite{Rui2019} on much larger scales, using $118$ galaxy-galaxy strong gravitational lensing systems, reports the constraint $\epsilon=-0.163\pm 0.001$ ($68\%$ CL) excluding GR. 

Here, we study the robust and accurate observational constraints on the Rastall gravity extension of the standard $\Lambda$CDM model. These would not only be important to see whether the Rastall gravity is a good candidate for studying the cosmological tensions discussed above or not, but also, as being robust and accurate, may provide a guide for the research community that studies the Rastall gravity in various aspects of gravitation and cosmology. Accordingly, we first construct the Rastall gravity extension of the base $\Lambda$CDM model at the background level (Sect.~\ref{sec:cosmo}) and elucidate its general relativistic equivalence (Sect.~\ref{sec:GRequiv}). We carry out a preliminary investigation of the model which provides a guide to its working and parameters (Sect.~\ref{sec:preliminary}). After deriving the linear perturbation equations (Sect.~\ref{sec:pert}), we constrain the model parameters using the latest full Planck CMB data, with and without including BAO data, in comparison to the standard $\Lambda$CDM model (Sect.~\ref{sec:obsanalysis}). We finally present a statistical comparison of the fit via Bayesian evidence (Sect.~\ref{sec:bayes}) and conclude our findings (Sect.~\ref{sec:Conclusions}).

\section{Rastall gravity extension of the base $\Lambda$CDM model}
\label{sec:cosmo} 
The Rastall gravity offers a simple generalization of the standard Einstein field equations of GR by relaxing the contribution of the term $g_{\mu\nu}R$ to the field equations and leads to the following modified Einstein field equations:
\begin{equation}
\label{eq:fieldeq}
R_{\mu\nu}-\left(\frac{1}{2}+\epsilon\right)g_{\mu\nu}R= \kappa  T_{\mu\nu},
\end{equation}
where $\kappa$ is Newton's constant scaled by a factor of $8\pi$ (and henceforth $\kappa=1$) and units are used such that $c=1$, $R_{\mu\nu}$ is the Ricci tensor, $R$ is the curvature scalar, $g_{\mu\nu}$ is the metric tensor, and $T_{\mu\nu}$ is the EMT describing the material content. The real constant $\epsilon$ is the Rastall parameter that measures the deviation from GR ($\epsilon=0$).

This modification in the spacetime geometry side (l.h.s.) of the Einstein field equations of GR corresponds to two simultaneous modifications of different nature in the material content side (r.h.s.): \textbf{(i)} Firstly, this modification is mathematically equivalent to adding, in a certain way, new contributions of the \textit{actual} material sources to the right hand side of the standard Einstein field equations, which then can be interpreted as an \textit{effective} source accompanying to the \textit{actual} material sources considered in the model. For, we can rewrite \eqref{eq:fieldeq} in a mathematically equivalent way as follows:
\begin{equation}
\label{eq:fieldeqT}
R_{\mu\nu}-\frac{1}{2}g_{\mu\nu}R=   T_{\mu\nu}+ \hat T_{\mu\nu},
\end{equation}
where
\begin{equation}
\label{eqn:EffectiveEMT}
\hat T_{\mu\nu}=-\frac{\epsilon}{1+4\epsilon}g_{\mu\nu}T
\end{equation}
is the EMT describing the effective source that arises from the actual material source. Here, we have made use of the relation $T\equiv g^{\mu\nu}T_{\mu\nu}=-(1+4\epsilon)R$ between the trace of the EMT of the actual material source and the curvature scalar, obtained by contracting \eqref{eq:fieldeq} with the inverse metric tensor $g^{\mu\nu}$. \textbf{(ii)} Secondly, this modification leads, in general, to a violation of the local conservation of the EMT of an actual material source (therefore, that of the effective source as well), as its divergence is not necessarily null, viz.,
\begin{equation}
\label{eq:cons_emt}
\nabla^{\mu}T_{\mu\nu}=-\epsilon\nabla_{\nu}R=\frac{\epsilon}{1+4\epsilon}\nabla_{\nu}T.
\end{equation}
It implies that the evolution of the energy density of an actual material source is, in general, modified compared to its usual evolution in general relativistic models. The reason is that only the covariant derivative of the Einstein tensor ($G_{\mu\nu}\equiv R_{\mu\nu}-\frac{1}{2}g_{\mu\nu}R$) part of the Rastall gravity \eqref{eq:fieldeq} is guaranteed to be null $\nabla^{\mu}G_{\mu\nu}=0$ through the twice-contracted Bianchi identity.

In this work, we study the Rastall gravity extension of the standard (the six-parameter base) $\Lambda$CDM model parameterized by only one additional degree of freedom, the Rastall parameter $\epsilon$, while keeping all the constituents (e.g., the physical ingredients of the Universe, the laws of the local physics) of the standard model as usual. Accordingly, we consider the spatially maximally symmetric and flat Friedmann-Robertson-Walker (FRW) spacetime
\begin{align}
\label{eq:metric}
{\rm d}s^2=-{\rm d}t^2+a^2(t)\,{\rm d}\vec{x}^2,
\end{align}  
where the scale factor $a(t)$ is function of proper time $t$ only. For describing the standard material content of the Universe, as usual, we consider the perfect fluid EMTs:
\begin{equation}
\label{eq:perfectEMT}
 T^{\mu\nu}_{i}=(\rho_i+p_i)u^{\mu}u^{\nu}+\,p_i\,g^{\mu\nu},
\end{equation}
where the index $i$ runs over the different actual sources described by the EoS of the form $p_i/\rho_i=w_i=\rm const.$ (with $\rho_i$ and $p_i$ being the energy density and the pressure of the $i^{\rm th}$ fluid, respectively), $u_\mu=(-1,0,0,0)$ is its velocity in the comoving frame (rest frame of the fluid) satisfying $u_{\mu}u^{\mu}=-1$ and $\nabla_{\nu}u^{\mu}u_{\mu}=0$.

We proceed with writing the modified Friedmann equations in a proper manner, namely, in a manner clearly identifying and handling the two simultaneous modifications of different nature in the material content side (r.h.s.) of the Einstein field equations of GR due to the Rastall gravity. We first note that the EMT describing the effective source is always of the form that of the usual vacuum energy of QFT:
\begin{equation}
\label{eq:effectiveEMT}
 \hat T^{\mu\nu}_{i}=\hat p_i\,g^{\mu\nu}.
\end{equation}
This could be deduced from
\begin{equation}
\label{eq:effectiverhop}
    \hat \rho_i=-\hat p_i=\frac{\epsilon}{1+4\epsilon}(3w_i-1)\rho_{i},
\end{equation}
which is obtained by using \eqref{eq:perfectEMT} in \eqref{eqn:EffectiveEMT} along with the trace of the EMT of the $i^{\rm th}$ fluid $T_i=\rho_{i}(3w_i-1)$.

The Einstein field equations of the model under consideration can explicitly be written as a set of two linearly independent differential equations with the unknown functions $H$ and $\rho_i$ as follows;
\begin{align}
\label{eq:rho}
3H^{2} =&\sum_i\rho_{i}+\rho_{\rm X},\\
-3H^{2}-2\dot{H}=&\sum_i w_i\rho_i+p_{\rm X},
\label{eq:pres}
\end{align}
where $H=\frac{\dot{a}}{a}$ is the Hubble parameter and the overdot denotes the derivative with respect to $t$. Note that $\rho_{\rm X}=\sum_i\hat \rho_i$ and $p_{\rm X}=\sum_i\hat p_i$ (satisfying $p_{\rm X}=-\rho_{\rm X}$) stand for the total energy density and pressure of the usual vacuum energy-like effective source accompanying to the actual sources and are not independent functions but are fully determined by the actual sources through \eqref{eq:effectiverhop}.

The dynamics of the actual sources---and hence the dynamics of the effective source as well---can directly be obtained from the continuity equation \eqref{eq:cons_emt}, which explicitly is
\begin{align}
\sum_i \big[\dot{\rho}_i+ 3H(1+ w_i)\rho_i\big]= \frac{\epsilon}{1+4\epsilon}\sum_i\dot{\rho}_i(1-3w_i).
\label{eq:contbackground}
\end{align}
It is reasonable to suppose that the fluids (actual sources) on cosmological scales are minimally interacting, i.e., interacting only gravitationally, which implies the separation of \eqref{eq:contbackground} for each type of fluid. In this case, we find the following redshift ($z=-1+1/a$) dependency for the background evolution of the energy density of the $i^{\rm th}$ fluid (actual source):
 \begin{align}
\rho_{i}=&\rho_{i0}\left(1+z\right)^{3(w_{i}+1)\left[1+\epsilon \frac{1-3w_{i}}{1+3\epsilon(1+w_{i})}\right]}.
\label{eq:rhoi}
\end{align}
Here and henceforth a subscript $0$ denotes the present-day ($z=0$) value of any quantity. We see that, except the cases $w_i=-1\,{\rm and}\,\frac{1}{3}$, the redshift dependence of the energy density of an actual source is modified with respect to its standard dependence $\rho_{i}=\rho_{i0} (1+z)^{3(1+w_i)}$. Thus, we have the usual relations $\rho_{{\rm vac}}=\rho_{{\rm vac0}}={\rm const.}$ for the usual vacuum energy $\left(w_{{\rm vac}}=-1\right)$ and $\rho_{{\rm r}}=\rho_{{\rm r0}}(1+z)^4$ for radiation $\left(w_{{\rm r}}=\frac{1}{3}\right)$, while a modified redshift dependence $\rho_{{\rm m}}=\rho_{{\rm m0}}(1+z)^{3+\frac{3\epsilon}{1+3\epsilon}}$ for dust $\left(w_{{\rm m}}=0\right)$. 

Next, considering \eqref{eq:effectiverhop} and \eqref{eq:rhoi}, it turns out that despite the fact that the effective source resembles the usual vacuum energy with an EoS parameter $\frac{p_{\rm X}}{\rho_{\rm X}}=w_{\rm X}=-1$, its energy density is not a constant, but
\begin{equation}
\label{eq:rhoX}
    \rho_{\rm X}=\frac{\epsilon}{1+4\epsilon}\sum_i (3w_i-1)\rho_{i0}\left(1+z\right)^{3(w_{i}+1)\left[1+\epsilon \frac{1-3w_{i}}{1+3\epsilon(1+w_{i})}\right]}.
\end{equation}
This obviously results from the non-conservation of the EMT of the actual material sources, see \eqref{eq:cons_emt}. Consequently, we have
\begin{equation}
\rho_{\rm X}=\hat\rho_{\rm vac}+\hat\rho_{\rm r}+\hat{\rho}_{\rm m},
\end{equation}
where
\begin{equation}
\begin{aligned}
\hat\rho_{\rm vac}=&-\frac{4\epsilon}{1+4\epsilon}\rho_{\rm vac0},\quad \hat\rho_{\rm r}=0\quad\textnormal{and}\quad \\ \hat\rho_{\rm m }=&-\frac{\epsilon}{1+4\epsilon}\rho_{\rm m0}\left(1+z\right)^{3+\frac{3\epsilon}{1+3\epsilon}}.
\label{eqn:hats}
\end{aligned}
\end{equation}
Note that $\hat\rho_{\rm r}=0$ due to the fact that radiation is traceless $\left(T=0\right)$. Therefore, it does not contribute to the effective source, i.e., to $\rho_{\rm X}$, as like it has been preserving its usual evolution $\rho_{\rm r}\propto(1+z)^{4}$. The usual vacuum energy ($w_{\rm vac}=-1$) of the QFT, on the other hand, still contributes to the Friedmann equation \eqref{eq:rho} like a cosmological constant, albeit with a rescaled value as $\rho_\Lambda=\rho_{\rm vac}+\hat\rho_{\rm vac}=\frac{1}{1+4\epsilon}\rho_{\rm vac}$. According to this, for a positive vacuum energy, $\rho_{\rm vac}>0$, to contribute to the Friedmann equation like a positive cosmological constant, $\rho_\Lambda>0$, there is a condition $\epsilon>-\frac{1}{4}$. In what follows, we stick to this condition considering the fact that the recent observations provide a clear detection of DE consistent with $\rho_{\Lambda}>0$ in the vicinity of present-day Universe, viz., for $z<1$ (see, for instance, \cite{Aubourg:2014yra}). The further condition $\epsilon>0$ leads to $\rho_{\rm X}<0$ (viz., $\hat\rho_{\rm vac}<0$ and $\hat\rho_{\rm m}<0$). Namely, under this condition, we have $\rho_{\rm X}(z=0)=-\frac{\epsilon}{1+4\epsilon}(4\rho_{\rm vac}+\rho_{\rm m0})<0$, and $\rho_{\rm X}$ continuously growing in larger negative values in the past. Thus, in this case ($\epsilon>0$), the effective source $\rho_{\rm X}$ dynamically screens the vacuum energy $\rho_{\rm vac}$ in the finite past, and in particular, the complete screening (viz., when $\rho_{\rm vac}+\rho_{\rm X}=0$) takes place at the redshift:
\begin{equation}
\label{eqn:zstar}
    z_*=\left(\frac{1}{\epsilon}\frac{\rho_{\rm vac0}}{\rho_{\rm m0}}\right)^{\frac{1+3\epsilon}{3+12\epsilon}}-1.
\end{equation}
This situation achieved for $\epsilon>0$ is of particular interest and tempting for its further theoretical and observational investigation, as it has recently been reported in \cite{Delubac:2014aqe,Aubourg:2014yra,Sahni:2014ooa,Mortsell:2018mfj,Poulin:2018zxs,Capozziello:2018jya,Wang:2018fng,Dutta:2018vmq,Banihashemi:2018oxo,Banihashemi:2018has,Visinelli:2019qqu,Akarsu:2019hmw,Ye:2020btb,Perez:2020cwa} that a number of persistent low-redshift tensions, including the $H_0$ tension, may be alleviated by a dynamical DE that assumes negative energy density values (or in cosmological models wherein the cosmological constant is dynamically screened) at finite redshift (see Sect. \ref{sec:intro}).

Finally, the modified Friedmann equation \eqref{eq:rho} for the Rastall gravity extension of the standard $\Lambda$CDM (Rastall-$\Lambda$CDM) reads
 \begin{equation}
\begin{aligned}
\label{eq:fried_1}
 3 H^2 =\rho_{\rm vac0}+\rho_{\rm m0}\left(1+z\right)^{3+\frac{3\epsilon}{1+3\epsilon}}+\rho_{\rm r0}\left(1+z\right)^{4}+\rho_{\rm X},
\end{aligned}
\end{equation}
where
\begin{equation}
    \rho_{\rm X}=-\frac{\epsilon}{1+4\epsilon}\left[4\rho_{\rm vac0}+\rho_{\rm m0}\left(1+z\right)^{3+\frac{3\epsilon}{1+3\epsilon}}\right].
\end{equation}
This can be rewritten in terms of density parameters, $\Omega_{i0}=\frac{\rho_{i0}}{3H_0^2}$, as follows:
\begin{equation}
\begin{aligned}
\label{eq:fried_3}
\frac{H^2}{H_0^2}=&\,\Omega_{\rm vac0}+\Omega_{\rm m0}\left(1+z\right)^{3+\frac{3\epsilon}{1+3\epsilon}}+\Omega_{\rm r0}\left(1+z\right)^4 \\
&+\Omega_{\rm X0}\frac{4\Omega_{\rm vac0}+\Omega_{\rm m0}\left(1+z\right)^{3+\frac{3\epsilon}{1+3\epsilon}}}{4\Omega_{\rm vac0}+\Omega_{\rm m0}},
\end{aligned}
\end{equation}
where
\begin{equation}\label{ox0}
\Omega_{\rm X0}=-\frac{\epsilon}{1+4\epsilon}\left(4\Omega_{\rm vac0}+\Omega_{\rm m0}\right),
\end{equation}
and the consistency relation $\Omega_{\rm vac0}+\Omega_{\rm m0}+\Omega_{\rm r0}+\Omega_{\rm X0}=1$ is satisfied.

\section{General relativistic equivalence: Non-minimally coupled $\Lambda$ and matter}
\label{sec:GRequiv}
As it was pointed out in \cite{Visser:2017gpz}, an equivalent of a cosmological model constructed within the Rastall gravity can always be constructed within the usual GR. But, this must be understood correctly. This does not imply that the Rastall gravity and GR are equivalent---namely, that the same set of sources leads to the same cosmological models---, but that the cosmological model in the presence of a certain set of sources constructed within the Rastall gravity can be reproduced upon a particular setup in the matter sector of the usual GR. When we compare the Einstein field equations of the usual GR, $R_{\mu\nu}-\frac{1}{2}g_{\mu\nu}R+\Lambda g_{\mu\nu}=\kappa T_{\mu\nu}$, to the Rastall gravity field equations \eqref{eq:fieldeq}, it turns out that the Rastall parameter $\epsilon$ sets the relation $\Lambda g_{\mu\nu}\rightarrow \epsilon R g_{\mu\nu}$ between the cosmological `constant' and scalar curvature---the two parameters that are of the same in nature---, which implies
\begin{align}
\label{eq:GRequiv}
\Lambda=-\epsilon R,
\end{align}
and then---through the twice-contracted Bianchi identity $\nabla^{\mu}(R_{\mu\nu}-\frac{1}{2}g_{\mu\nu}R)=0$ leading to the relation $\nabla^{\mu}(\kappa T_{\mu\nu}+g_{\mu\nu}\Lambda)=0$---a non-minimal interaction between the cosmological `constant' and matter fields as
\begin{align}
\label{eq:transfer}
\nabla^{\mu}T_{\mu\nu}=Qu_{\nu}\quad\textnormal{and}\quad \nabla_{\nu}\Lambda=-Qu_{\nu},
\end{align}
described by the energy-momentum transfer function
\begin{align}
\label{eq:emtrans}
Qu_{\nu}=\frac{\epsilon}{1+4\epsilon}\nabla_{\nu}T.
\end{align}
Thus, the Rastall gravity extension of the standard $\Lambda$CDM (Rastall-$\Lambda$CDM) model under consideration in this study, corresponds to a very particular cosmological setup within the usual GR, namely, to a uniquely tailored extension of the standard $\Lambda$CDM model, wherein the cosmological `constant' now is a dynamical quantity---relying on its non-minimal interaction with the matter fields---in a particular way that guarantees it to remain proportional to the curvature scalar. Namely, in principle, we could have constructed exactly the same cosmological model under consideration here well within the usual GR from the beginning. Indeed, one may check that it is equivalent to a general relativistic cosmological model wherein the energy transfer function between the cosmological `constant' and matter fields is given in the following cumbersome form:
\begin{equation}
\begin{aligned}
\label{eq:Qdef}
Q=&\frac{\epsilon}{1+4\epsilon}H(z)(1+z)\sum_i (1-3w_i)\frac{{\rm d}\rho_{i}}{{\rm d}z} \\
=&\frac{\sqrt{3}\epsilon}{1+3\epsilon}\rho_{\rm m0}\left(1+z\right)^{3+\frac{3\epsilon}{1+3\epsilon}}\\ 
&\times\bigg[\frac{1}{1+4\epsilon}
\rho_{\rm vac0}+\frac{1+3\epsilon}{1+4\epsilon}\rho_{\rm m0}\left(1+z\right)^{3+\frac{3\epsilon}{1+3\epsilon}} \\
&\,\,\quad +\rho_{\rm r0}\left(1+z\right)^{4}\bigg]^{1/2}.
\end{aligned}
\end{equation}
Indeed, it is obvious that the Rastall-$\Lambda$CDM model---wherein we simply control the coupling strength of the curvature scalar to the gravity via the Rastall parameter $\epsilon$---is completely different than the standard $\Lambda$CDM  model, and moreover the alternative construction of this model within the usual GR from the beginning, is neither trivial nor economical. Yet we have learned one another important lesson from this discussion: The model under consideration here presents a particular example of general relativistic cosmological models wherein there is non-minimal interaction within the dark sector \cite{Barrow:2006hia,Wang:2016lxa}---i.e., between the `cold' dark matter and DE (in our case $\Lambda$). Such cosmological models are of particular interest in the current cosmological studies \cite{Kumar:2016zpg,Yang:2017zjs,DiValentino:2017iww,Yang:2018euj,Kumar:2019wfs,DiValentino:2019ffd,Yang:2019uog} as it has recently turned out that they are potential candidates for addressing the so-called $H_0$ tension prevailing within the standard $\Lambda$CDM model. Therefore, it is conceivable that the Rastall-$\Lambda$CDM model must also be a potential candidate for addressing the $H_0$ tension. Indeed, we will show in what follows, without switching to the GR construction of the model, that the deviation term in the Rastall gravity, $\epsilon g_{\mu\nu}R$, can dynamically screen the usual vacuum energy at high redshifts, which in turn leads to larger $H_0$ values compared to the standard $\Lambda$CDM.

\section{A preliminary investigation} 
\label{sec:preliminary}
In this section, we present a preliminary investigation for a demonstration of how the Rastall-$\Lambda$CDM model works, and a guide to the values of the parameters of it. We first derive some useful parameters that we shall use to discuss some features/limitations of the model.

First of all, since the usual radiation evolution is not affected from the Rastall gravity extension, we can safely use the relevant standard equations. The photon energy density today $\rho_{{\gamma}0}$ is then still well constrained, relying on a simple  relation: $\rho_{\gamma}=\frac{\pi^2}{15} T_{\rm CMB}^4$ with the CMB monopole temperature  \cite{Dodelson03}, which is very precisely measured to be $T_{{\rm CMB}0}=2.7255\pm 0.0006\,{\rm K}$ \cite{Fixsen09}. We suppose, in line with standard particle physics, three neutrino species ($N_{\rm eff}=3.046$) with minimum allowed mass $\sum m_{\nu}=0.06\, {\rm eV}$. Then, the radiation density parameter can be given in the standard way: $\Omega_{{\rm r}0}=\Omega_{{\rm \gamma}0}+\Omega_{{\nu}0}=4.18343\times 10^{-5} h^{-2}$, where $h$ is the dimensionless reduced Hubble constant parametrizing the Hubble constant via $H_0=100\,h\, {\rm km\,s}^{-1}{\rm Mpc}^{-1}$ \cite{Dodelson03}. Using a reasonable value, for instance, $H_0=68\, {\rm km\,s}^{-1}{\rm Mpc}^{-1}$, we find that the density parameter of the radiation today is negligible, viz., $\Omega_{{\rm r}0}=9.0472 \times 10^{-5}$. Neglecting this small contribution of radiation today, we have $\Omega_{\rm vac0}+\Omega_{\rm m0}+\Omega_{\rm X0}=1$. Using this relation with \eqref{ox0}, we can write the present-day density parameter of the effective source as follows:
\begin{equation}
    \Omega_{\rm X0}=-4\epsilon+3\,\epsilon\,\Omega_{\rm m0}.
    \label{eq:woutrad}
\end{equation}
From this, we read off $\frac{\rho_{\rm X0}}{\rho_{\rm m0}}=-\epsilon\,\left(\frac{4}{\Omega_{\rm m0}}-3\right)$, while we have $\frac{\rho_{\rm X}}{\rho_{\rm m}}\approx-\frac{\epsilon}{1+4\epsilon}$ for $z\gg0$, say, in the early Universe, see \eqref{eq:rhoi} and \eqref{eqn:hats}. Using the relation \eqref{eq:woutrad} along with \eqref{eqn:zstar}, the redshift at which the vacuum energy is completely screened by the effective source reads as
\begin{equation}
    z_*=\left(\frac{1}{\epsilon\,\Omega_{\rm m0}}+\frac{4}{\Omega_{\rm m0}}-\frac{1}{\epsilon}-3\right)^{\frac{1+3\epsilon}{3+12\epsilon}}-1.
    \label{eq:zstarreduced}
\end{equation}

Since the evolution of radiation remains unaltered in the Rastall gravity, we basically do not expect any modification in the standard history of the Universe throughout the radiation epoch. Yet, the modified evolution of dust would have consequences on the transition from radiation to dust domination. The radiation-matter (dust) transition is one of the most important epochs in the history of the Universe, as it alters the growth rate of density perturbations: during the radiation epoch, which yields $H^2(z)\propto (1+z)^{4}$, perturbations well inside the horizon are nearly frozen but once matter domination commences as $H(z)$ flattens to yield $H^2(z)\propto (1+z)^{3}$ during the dust epoch, perturbations on all length scales are able to grow by gravitational instability and therefore it sets the maximum of the matter power spectrum. The modified matter-radiation equality ($\rho_{\rm r}=\rho_{\rm m}$) redshift reads
\begin{equation}
\begin{aligned}
\label{eq:keq}
z_{\rm eq}=\left(\frac{\Omega_{\rm m0}}{\Omega_{\rm r0}}\right)^{1+3\epsilon}-1,
\end{aligned}
\end{equation}
where, for a given value of the ratio of the density parameters---of course, we suppose $\frac{\Omega_{\rm m0}}{\Omega_{\rm r0}}>1$---positive (negative) $\epsilon$ values shift $z_{\rm eq}$ to larger (smaller) values. This in turn shifts the turnover in the matter power spectrum via a highly sensitive parameter to the modifications to GR, namely, the wavenumber of a mode that enters the horizon at the radiation-matter transition:
\begin{equation}
\begin{aligned}
\label{eq:keq}
k_{\rm eq}=\frac{H_{\rm eq}}{1+z_{\rm eq}}=H_0 \sqrt{\frac{2+7\epsilon}{1+4\epsilon}\Omega_{\rm m0}\left(\frac{\Omega_{\rm m0}}{\Omega_{\rm r0}}\right)^{1+6\epsilon}},
\end{aligned}
\end{equation}
where we have ignored $\Omega_{{\rm vac}0}$ since its contribution is safely negligible.

Note that the condition $\epsilon>-\frac{1}{4}$ that we introduced for $\rho_{\Lambda}>0$ in the previous section, ensures the real positive values of $k_{\rm eq}$ and $H^2(z)\propto (1+z)^{3+\frac{3\epsilon}{1+3\epsilon}}$ during the dust era to be flatter than $H^2(z)\propto (1+z)^{4}$ during the radiation era. These two parameters, $H_{\rm eq}$ and $k_{\rm eq}$, are not expected to deviate much from the ones obtained within the standard $\Lambda$CDM model. Therefore, these are very useful to give an opinion whether a cosmological model is well behaved at high redshifts, for instance, with regard to the CMB data relevant to $z\sim1100$.

We can now make use of the parameters derived here for a preliminary investigation of the model: The values of these parameters may be utilized for making estimations on $\epsilon$ by manipulating the late time dynamics of the Universe in our model, for instance, to better describe the existing model independent $H_0$ data as well as the BAO data from $z\lesssim2.4$. We, of course, must also check the price paid for this manipulation from the chosen $\epsilon$ value in the dynamics of the earlier Universe, for instance, in the cosmological parameters physically related/sensitive to the presence/amount of radiation.

\begin{figure}[t!]\centering
 \includegraphics[width=0.38\textwidth]{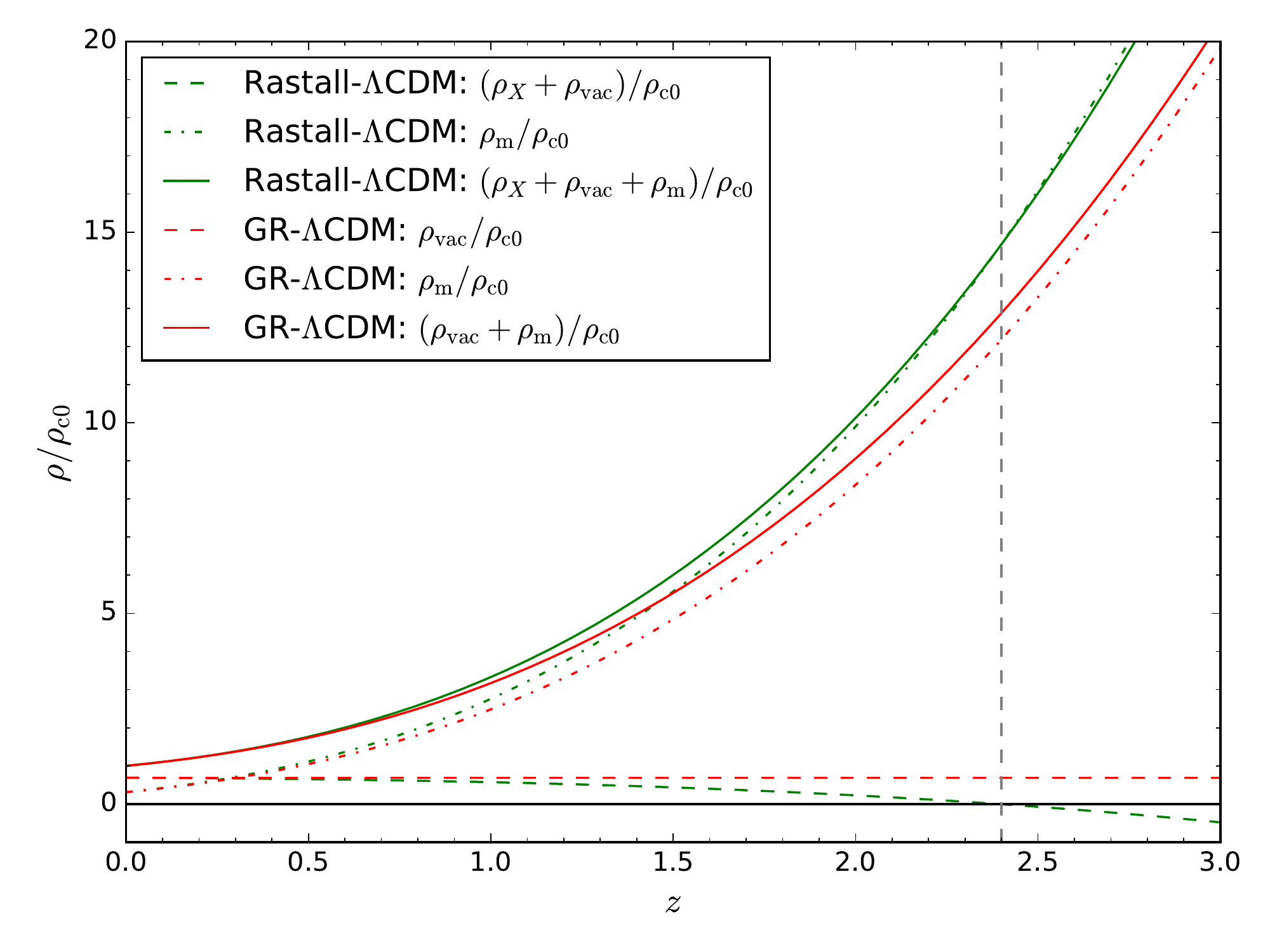} 
 \caption{Plot of $\rho/\rho_{\rm c0}$ vs $z$ for different combinations of energy densities. In all cases, we use $\Omega_{\rm m0}= 0.31$ and $H_0=68\, {\rm km\,s}^{-1}{\rm Mpc}^{-1}$ and $\epsilon=0.06$. Here the vertical dashed line refers to the redshift $z_*=2.4$, at which the effective source $\rho_{\rm X}$ completely screens the usual vacuum energy $\rho_{\rm vac}$.} 
 \label{fig:rhoxc}
  \end{figure}
  
We proceed with a reasonable set of values: $\Omega_{\rm m0}= 0.31$ and $H_0=68\, {\rm km\,s}^{-1}{\rm Mpc}^{-1}$, acceptable with regard to both the recent Planck predictions \cite{Aghanim:2018eyx} and the model independent tip of the red giant branch (TRGB) $H_0$ value \cite{Freedman:2019jwv}, along with the choice of $\epsilon=0.06$. We find, for the present-day Universe, a considerable amount of negative contribution of the effective source, $\Omega_{\rm X0}=-0.184$ [see \eqref{eq:woutrad}], which is then compensated by the increased value of the usual vacuum energy, $\Omega_{\rm vac0}=0.874$, so that to yield a cosmological constant-like contribution, $\Omega_{\Lambda0}=0.69$, as in the standard $\Lambda$CDM ($\epsilon=0$). On the other hand, this enhanced amount of the usual vacuum is completely screened by the effective source at $z_*=2.4$ [see \eqref{eq:zstarreduced} and Figure \ref{fig:rhoxc}], which is pretty close to the values suggested in \cite{Aubourg:2014yra,Sahni:2014ooa,Akarsu:2019hmw} for relaxing a number of persistent low-redshift tensions,  including  the $H_0$ tension, that arise within the standard $\Lambda$CDM model. It is worth noting that, as may be seen in the same figure, for $z_*>2.4$, the dust energy density assumes values larger than the total energy density of the Universe. Yet, the ratio of the energy density of the effective source to that of the dust is $\frac{\rho_{\rm X0}}{\rho_{\rm m0}}=-0.59$ today but it settles in a value pretty close to zero, $\frac{\rho_{\rm X}}{\rho_{\rm m}}\approx-0.048$, at high redshifts ($z\gg0$). This implies that the impact of the effective source on the dynamics of the Universe diminishes (yet not completely) with the increasing redshift. However, the price (due to the modification in the EMT conservation) we paid for this tempting result (say, the screening of the vacuum energy) is that the dust energy density grows considerably faster than it does in the usual GR, $\rho_{\rm m}\propto (1+z)^{3.15}$ (which is also tracked by the effective source at high redshifts), whereas the radiation energy density always grows as usual, $\rho_{\rm r}\propto (1+z)^{4}$. This leads to unrealistic values for the key parameters relevant to the early Universe, namely, $z_{\rm eq}=15303$ and $k_{\rm eq}=0.04470 \, {\rm Mpc}^{-1}$, which are extremely different than the values $z_{\rm eq}=3391$ and $k_{\rm eq}=0.01045\, {\rm Mpc}^{-1}$ obtained in the case of the standard $\Lambda$CDM model ($\epsilon =0$). This situation signals that the Rastall-$\Lambda$CDM model with $\epsilon=0.06$, which is tempting as it leads to $z_*=2.4$ in line with \cite{Aubourg:2014yra,Sahni:2014ooa,Akarsu:2019hmw}, is not well behaved at large redshifts. Thus, it is conceivable that the high redshift cosmological data would not allow such large positive values of $\epsilon$.
  
  \begin{figure}[t!]\centering
 \includegraphics[width=0.37\textwidth]{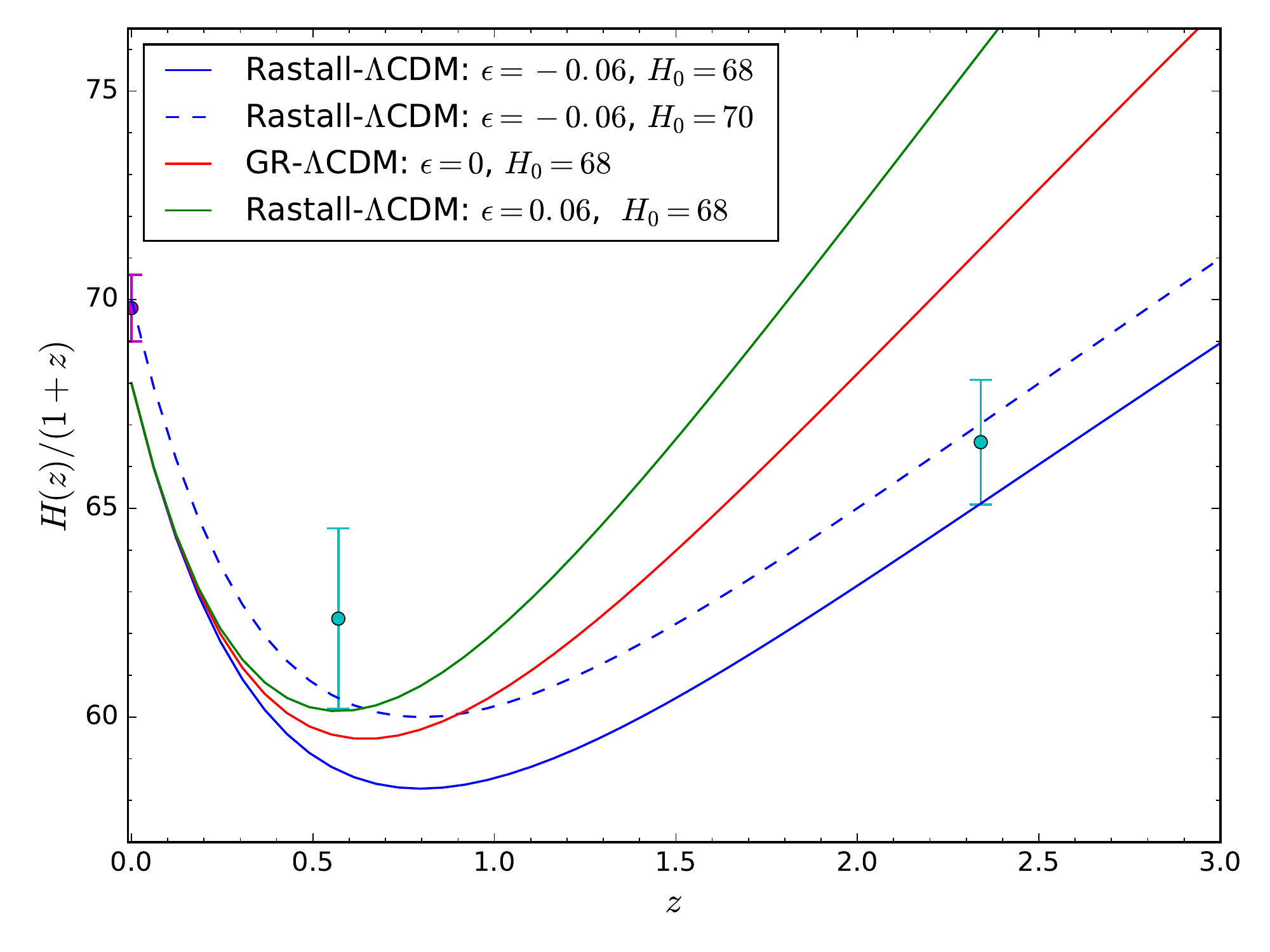} 
 \caption{Plot of $H(z)/(1+z)$ vs $z$ for some selected values of $\epsilon$ and $H_0$ as shown in the legend. In all cases, we use $\Omega_{{\rm m}0}=0.31$. The three error bars stand for  $H_0= 69.8 \pm 0.8\, {\rm km\,s}^{-1}{\rm Mpc}^{-1}$ from the TRGB $H_0$ \cite{Freedman:2019jwv}, $H(z=0.57) = 97.9 \pm 3.4\,{\rm km\,s}^{-1}{\rm Mpc}^{-1}$ \cite{bao3}, and $H(z=2.34) = 222.4 \pm 5.0\,{\rm km\,s}^{-1}{\rm Mpc}^{-1}$ from the latest BAO data \cite{Delubac:2014aqe}.} 
 \label{fig:hz}
  \end{figure} 

Next, we proceed to have a closer look at the dynamics of the Universe by focusing on a narrow redshift range $0\leq z\leq 3$, within which we can, in a relatively straightforward way, compare the $H(z)$ function of the Rastall-$\Lambda$CDM model with the model independent $H_0$ measurements, e.g., $H_0= 69.8 \pm 0.8\, {\rm km\,s}^{-1}{\rm Mpc}^{-1}$ from the TRGB $H_0$ \cite{Freedman:2019jwv}, and the latest high precision BAO data: $H(z=0.57) = 97.9 \pm 3.4\,{\rm km\,s}^{-1}{\rm Mpc}^{-1}$ \cite{bao3} and $H(z=2.34) = 222.4 \pm 5.0\,{\rm km\,s}^{-1}{\rm Mpc}^{-1}$ \cite{Delubac:2014aqe}. To do so, along with these data points, we plot the function $H(z)/(1+z)$ versus $z$ in Figure \ref{fig:hz} considering four different pair of values of the parameters $\epsilon$ and $H_0$ as mentioned in the legend of this figure, wherein we keep $\Omega_{\rm m0}=0.31$ in all the cases (two of which correspond to the cases in Figure \ref{fig:rhoxc}). First, we notice that the Rastall-$\Lambda$CDM model (green curve, $\epsilon=0.06$, $H_0= 68\, {\rm km\,s}^{-1}{\rm Mpc}^{-1}$, which leads to $z_*=2.4$) does better than the GR-$\Lambda$CDM (red curve, $\epsilon=0$, $H_0= 68\, {\rm km\,s}^{-1}{\rm Mpc}^{-1}$) to describe the lower-redshift BAO data ($z=0.57$), but the faster growth of  $H(z)/(1+z)$ at high redshifts leads to an increased tension of the Rastall-$\Lambda$CDM model with the high-redshift BAO Ly-$\alpha$ data ($z=2.34$) compared to the standard $\Lambda$CDM model. This seems to suggest that negative values of $\epsilon$ can help better representation of the high-redshift BAO Ly-$\alpha$ data ($z=2.34$) in the Rastall-$\Lambda$CDM model, which of course implies compromise from the feature of screening the usual vacuum energy by the effective source at a finite redshift. For, we see that the Rastall-$\Lambda$CDM model (blue solid curve,  $\epsilon=-0.06$, $H_0= 68\, {\rm km\,s}^{-1}{\rm Mpc}^{-1}$) better represents the high-redshift BAO Ly-$\alpha$ data ($z=2.34$) due to slow growth of $H(z)/(1+z)$ at higher redshifts. But in this case, this model worsens in representing the lower-redshift BAO data ($z=0.57$) compared to the GR-$\Lambda$CDM model.  However, surprisingly, if we use a larger $H_0$ value as well, for instance,  $H_0= 70\, {\rm km\,s}^{-1}{\rm Mpc}^{-1}$ very close to the model independent TRGB $H_0$ measurement, it turns out that the Rastall-$\Lambda$CDM model (blue dashed curve,  $\epsilon=-0.06$, $H_0= 70\, {\rm km\,s}^{-1}{\rm Mpc}^{-1}$) reconciles with all the three data points simultaneously. This makes negative $\epsilon$ values promising with regard to addressing the so called $H_0$ tension on top of a good description of the both BAO data (the standard $\Lambda$CDM model has known to be suffering from). However, most likely, it would lead to an inconsistency with the CMB data, as the values of $\epsilon$ leading to a significant improvement in this direction would give rise to unacceptable amount of shifts in the values of $z_{\rm eq}$ and $k_{\rm eq}$. Indeed, $\epsilon=-0.06$, that we have used just to develop an opinion, leads to the unacceptable values $z_{\rm eq}=790$ and $k_{\rm eq}\sim0.025 \,{\rm Mpc}^{-1}$, which obviously signals spoiling of a successful description of the early Universe.

The lesson we learned in this section may be summarized as follows: Through this preliminary investigation, it is not possible to reach  to a decisive conclusion whether the $H_0$ and/or BAO data show tendency of $\epsilon$ deviating from zero (GR) in a certain direction. Moreover, a significant improvement with regard to $H_0$ and/or BAO data would most likely lead to spoiling of a successful description of the early Universe, which signals that CMB data would keep $\epsilon$ values close to zero. Therefore, we expect only an insignificant deviation from the standard $\Lambda$CDM model when it is extended via the Rastall gravity. A conclusive answer, of course, cannot be given unless we rigorously confront the model with the observational data.

\section{Linear perturbations}
\label{sec:pert} 

In this section, we derive the general form of the equations which describe small cosmological perturbations within the Rastall-$\Lambda$CDM model. We consider the perturbed RW metric, $g_{\mu\nu}= g_{\mu\nu}^{(0)} + h_{\mu\nu}$, where $g_{\mu\nu}^{(0)}$ indicates the background of the spatially flat RW metric \eqref{eq:metric} with a small fluctuation, $h_{\mu\nu}$, and we choose the synchronous gauge ($h_{\mu0} = 0$). The line element has the form
\begin{equation}
\begin{aligned}
\label{pertrw}
{\rm d}s^2= a(\eta)^2\left[-{\rm d}\eta^2+\left(\delta_{jk}+h_{jk}\right) {\rm d}x^{j}{\rm d}x^{k}\right],
\end{aligned}
\end{equation}
where $x^{j}$, with $j=(1,2,3)$, are the spatial components in Cartesian coordinates and $\eta$ is the conformal time. The comoving coordinates are related to the proper time $t$ and positions $r$ by ${\rm d}x^0={\rm d}\eta=\frac{{\rm d}t}{a(\eta)}$, ${\rm d}x=\frac{{\rm d}r}{a(\eta)}$. We introduce the perturbations as follows:
\begin{eqnarray}
\rho_i = \rho_i^{(0)}+\delta\rho_i, \;\; p_{i}= p_{i}^{(0)}+\delta p_{i},\;\; u_{i}=u_{i}^{(0)}+\delta u_{i},\quad \;\;
\end{eqnarray}
where the superscript $(0)$ indicates the background functions and $\delta\rho_{i}$, $\delta u_{i}$ and  $\delta p_{i}$ are the perturbed quantities in energy density, four-velocity and pressure, respectively. We also introduce the following definitions:
\begin{align}
\delta_{i} \equiv\frac{\delta \rho_{i}}{\rho_{i}} \,,\quad c_{s,i}^2=\frac{\delta p_i}{\delta \rho_i} \,,\quad \theta \equiv \partial_{k}\delta u^{k}\,, \quad h \equiv \frac{h^{k}{}_k}{a^{2}},
\end{align}
where $c_{s,i}^2$ is adiabatic sound speed squared.

Within the formalism, the continuity equation reads
\begin{equation}
\begin{aligned}
\label{eq:cont}
\sum_i \Bigg[&\delta'_{i}\left[ \frac{1+3\epsilon(1+w_i)}{1+4\epsilon}\right]+3\mathcal{H}\left(c_{s,i}^2-w_i\right)\delta_i\\
&+(1+w_i)\left(\theta_{i}+\frac{h'}{2}\right) \Bigg]=0,
\end{aligned}
\end{equation}
from the perturbations of the actual EMT conservation equation [see  \eqref{eq:cons_emt}] for $\nu=0$. Here the prime denotes derivative with respect to $\eta$ and $\mathcal{H}=\frac{a'}{a}$. For $\nu=i$, Euler equation reads:
\begin{equation}
\begin{aligned}
\label{eq:euler}
\sum_i \Bigg[&\theta'_{i}+ \frac{1-3w_i}{1+3\epsilon(1+w_i)}\mathcal{H}\theta_{i}\\
&-\frac{k^2\delta_i}{(1+4\epsilon)(1+w_i)}\left[\epsilon+(1+\epsilon)c_{s,i}^2\right]\Bigg]=0.
\end{aligned}
\end{equation}
The relativistic species remain unaltered in Rastall gravity both at the background and perturbative levels. Thus, the Boltzmann hierarchy for the relativistic relics follows the standard procedures as described in \cite{Ma_Bertschinger} (see also \cite{class}). We suppose that the evolution of baryons ($w_{\rm b}=c_{s,\rm b}^2=0$) is not altered by the Rastall gravity but that of the cold dark matter ($w_{\rm cdm}=c_{s,\rm cdm}^2=0$), for which we modify the procedures given in \cite{Ma_Bertschinger,class} accordingly \cite{footnote}. And, the first order continuity and Euler equations from \eqref{eq:cont} and \eqref{eq:euler} read
\begin{align}
\label{eq:contdust}
&\delta'_{\rm cdm}\frac{1+3\epsilon}{1+4\epsilon}+\theta_{\rm cdm}+\frac{h'}{2}=0, \\
&\theta'_{\rm cdm}+\frac{1}{1+3\epsilon}\mathcal{H}\theta_{\rm cdm}-\frac{\epsilon}{1+4\epsilon}k^2\delta_{\rm cdm}=0.
\label{eq:eulerdust}
\end{align}

\section{Observational constraints}
\label{sec:obsanalysis}

Considering the background and perturbation dynamics presented above, in what follows, we explore the full parameter space of the Rastall-$\Lambda$CDM model---namely, the Rastall gravity extension of the six-parameter base $\Lambda$CDM based on GR via the Rastall parameter $\epsilon$---, and, for comparison, that of the standard $\Lambda$CDM (GR-$\Lambda$CDM) model. The baseline seven free parameters set of the Rastall-$\Lambda$CDM model is, therefore:
\begin{equation*}
\label{baseline1}
\mathcal{P}= \Big\{ \omega_{\rm b}, \, \omega_{\rm cdm}, \, \theta_s, \,  A_s, \, n_s, \, \tau_{\rm reio}, 
\,   \epsilon \Big\},
\end{equation*}
where the first six parameters are the baseline parameters of the standard $\Lambda$CDM model, namely: $\omega_{\rm b}$ and $\omega_{\rm cdm}$ are respectively the dimensionless densities of baryons and cold dark matter; $\theta_s$ is the ratio of the sound horizon to the angular diameter distance at decoupling; $A_s$ and $n_s$ are respectively the amplitude and spectral index of the primordial curvature perturbations, and $\tau_{\rm reio}$ is the optical depth to reionization. The uniform priors used for the model parameters are $\omega_b\in[0.018,0.024]$, $\omega_{\rm cdm}\in[0.10,0.14]$, $100\,\theta_{s}\in[1.03,1.05]$, $\ln(10^{10}A_s)\in[3.0,3.18]$, $n_s\in[0.9,1.1]$, $\tau_{\rm reio}\in[0.04,0.125]$, and $\epsilon\in[-0.04,0.04]$.

In order to constrain the models, we use the latest Planck CMB and BAO data: We use the recently released full Planck-2018 \cite{Aghanim:2018eyx} CMB temperature and polarization data which comprise of the low-$l$ temperature and polarization likelihoods at $l \leq 29$, temperature (TT) at $l \geq 30$, polarization (EE) power spectra, and cross correlation of temperature and  polarization (TE). The Planck-2018 CMB lensing power spectrum likelihood \cite{Planck2018:GL} is also included. Along with the Planck CMB data, we consider the measurements of BAO provided by the distribution of galaxies in galaxy-redshift surveys. We use BAO distance measurements probed by (i) Six  Degree  Field  Galaxy  Survey  (6dFGS) at effective redshift $z_{\rm eff} = 0.106$ \cite{bao1}, (ii) 
the  Main  Galaxy  Sample  of  Data  Release 7  of  Sloan  Digital  Sky  Survey  (SDSS-MGS) at effective redshift $z_{\rm eff} = 0.15$ \cite{bao2}, (iii) 
the  LOWZ  and  CMASS  galaxy  samples  of Data Release 11 (DR11) of   the Baryon  Oscillation  Spectroscopic  Survey  (BOSS) LOWZ  and  BOSS-CMASS at effective redshifts $z_{\rm eff} = 0.32$ and $z_{\rm eff} = 0.57$, respectively \cite{bao3}, (iv) correlation of Lyman-$\alpha$ forest absorption and quasars at $z_{\rm eff} = 2.35$ obtained in SDSS DR14 \cite{bao4}. Also, we use the measurement obtained in \cite{bao5}, where the BAO scale is measured at $z_{\rm eff} =2.34$.

\begin{table}[b!]
\small
\caption{Constraints (68\% and 95\% CLs)  on  the free and some derived parameters of the Rastall-$\Lambda$CDM and GR-$\Lambda$CDM models for CMB and CMB+BAO data. The parameter $H_{\rm 0}$ is measured in the units of km s${}^{-1}$ Mpc${}^{-1}$. The entries in blue color represent the constraints on the corresponding GR-$\Lambda$CDM parameters.}
\label{tableI}
\setlength\extrarowheight{2pt}
\begin{tabular} { |l| l| l| l| l|  }  \hline 
 Parameter &  CMB     & CMB + BAO        \\ 
\hline
$10^{2}\omega_{\rm b }$ &   $2.246^{+ 0.016+0.031}_{-0.016-0.031}$  &  $2.238^{+ 0.014+0.028}_{-0.014-0.028}$  \\[1ex]

& \textcolor{blue}{$2.236^{+0.013+0.026}_{-0.013-0.026}$}    &  \textcolor{blue}{$2.245^{+ 0.012+0.025}_{-0.012-0.024}$}\\[1ex]
 \hline
 $\omega_{\rm cdm }$  & $0.1235^{+ 0.0028+0.0055}_{-0.0028-0.0055}$  &  $0.1185^{+ 0.0012+0.0023}_{-0.0012-0.0023}$   \\[1ex]
 
         & \textcolor{blue}{$0.1202^{+0.0012+0.0023}_{-0.0012-0.0022}$}   &  \textcolor{blue}{$0.1189^{+0.0010+0.0019}_{-0.0010-0.0018}$}  \\[1ex]
\hline

$100 \theta_{s } $ &  $1.0418^{+0.0003+0.0006}_{-0.0003-0.0006}$ &  $1.0420^{+0.0003+0.0006}_{-0.0003-0.0005}$ \\[1ex]

&  \textcolor{blue}{ $1.0419^{ +0.0003+0.0006}_{-0.0003-0.0006}$ }   & \textcolor{blue}{$1.0420^{ +0.0003+0.0006}_{-0.0003-0.0006} $}    \\[1ex]
\hline
$\ln10^{10}A_{s }$ &   $ 3.042^{+0.011+0.025}_{-0.013-0.026}   $ & $3.050^{+0.013+0.028}_{-0.015-0.025}  $ \\[1ex]

&  \textcolor{blue}{$3.046^{+0.012+0.026}_{-0.014-0.025}$}   & \textcolor{blue}{$3.051^{+ 0.013+0.027}_{-0.014-0.026}$}   \\[1ex]
\hline
$n_{s } $ &  $0.961^{+0.004+0.008}_{-0.004-0.008} $  &  $0.967^{+0.003+0.006}_{-0.003-0.006}  $  \\[1ex]

&  \textcolor{blue}{$0.964^{+0.003+0.007}_{-0.003-0.007}$}  & \textcolor{blue}{ $0.967^{+0.003+0.006}_{-0.003-0.006}  $}    \\[1ex]
\hline
$\tau_{\rm reio } $  &  $0.053^{+0.006+0.013}_{-0.007-0.012}   $ &$0.058^{+0.007+0.014}_{-0.008-0.014}   $\\[1ex]

& \textcolor{blue}{$0.055^{+0.006+0.014}_{-0.008-0.013}$}   &  \textcolor{blue}{$0.059^{+ 0.007+0.014}_{-0.007-0.014}$} \\[1ex]
\hline
  $\epsilon$  & $-0.0010^{+0.0008+0.0015}_{-0.0008-0.0015}$      &  $ 0.0003^{+0.0004+0.0008}_{-0.0004-0.0008}$
            \\[1ex]
          & \textcolor{blue}{$[0]$}  &  \textcolor{blue}{$[0]$}\\ [1ex]

\hline
\hline

$\Omega_{\rm{m0} }$   &  $0.347^{+0.024+0.054}_{-0.027-0.047}   $  &  $0.304^{+ 0.009+0.019}_{-0.009-0.017}   $ \\[1ex]

& \textcolor{blue}{$0.316^{+ 0.007+0.014}_{-0.007-0.013}$}  & \textcolor{blue}{$0.308^{+ 0.006+0.011}_{-0.006-0.011}$}    \\[1ex]

\hline

$\Omega_{\rm{vac0} }$   &  $0.650^{+0.029+0.052}_{-0.026-0.058}$  &  $0.697^{+0.010+0.020}_{-0.010-0.021}    $ \\[1ex]

& \textcolor{blue}{$0.684^{+ 0.007+0.013}_{-0.007-0.014}$}  & \textcolor{blue}{$0.692^{+ 0.006+0.011}_{-0.006-0.011}$}    \\[1ex]

\hline

$\Omega_{\rm{X0} }$   &  $0.0030^{+0.0023+0.0045}_{-0.0023-0.0046} $  &  $-0.0010^{+0.0013+0.0026}_{-0.0013-0.0026}   $ \\[1ex]

& \textcolor{blue}{$[0]$}  & \textcolor{blue}{$[0]$}    \\[1ex]

\hline

$\Omega_{\Lambda0 }$   &  $ 0.653^{+0.028+0.048}_{-0.024-0.055}$  &  $0.696^{+0.009+0.018}_{-0.009-0.019}$ \\[1ex]

& \textcolor{blue}{$0.684^{+ 0.007+0.013}_{-0.007-0.014}$}  & \textcolor{blue}{$0.692^{+ 0.006+0.011}_{-0.006-0.011}$}      \\[1ex]

\hline
 
$H_{\rm 0}$ &   $65.10^{+1.80+3.50}_{-1.80-3.50} $ &  $ 68.31^{+ 0.76+1.50}_{-0.76-1.50}  $ 
 \\[1ex]
 
 & \textcolor{blue}{$67.30^{+0.51+0.98}_{-0.51-0.96} $}  & \textcolor{blue}{$67.92^{+0.43+0.83}_{-0.43-0.82}$} \\[1ex]
 \hline
  $\sigma_{8}$ &  $0.792^{+0.017+0.035}_{-0.017-0.032}  $ &  $0.818^{+ 0.011+0.021}_{-0.011-0.022} $ \\[1ex]
  
  & \textcolor{blue}{$0.812^{+ 0.006+0.012}_{-0.006-0.011}$}  & \textcolor{blue}{$0.810^{+0.006+0.012}_{-0.006-0.011}$}  \\[1ex]
  \hline

\end{tabular}
\end{table}

\begin{figure*}[htt!]
\includegraphics[width=14.5cm]{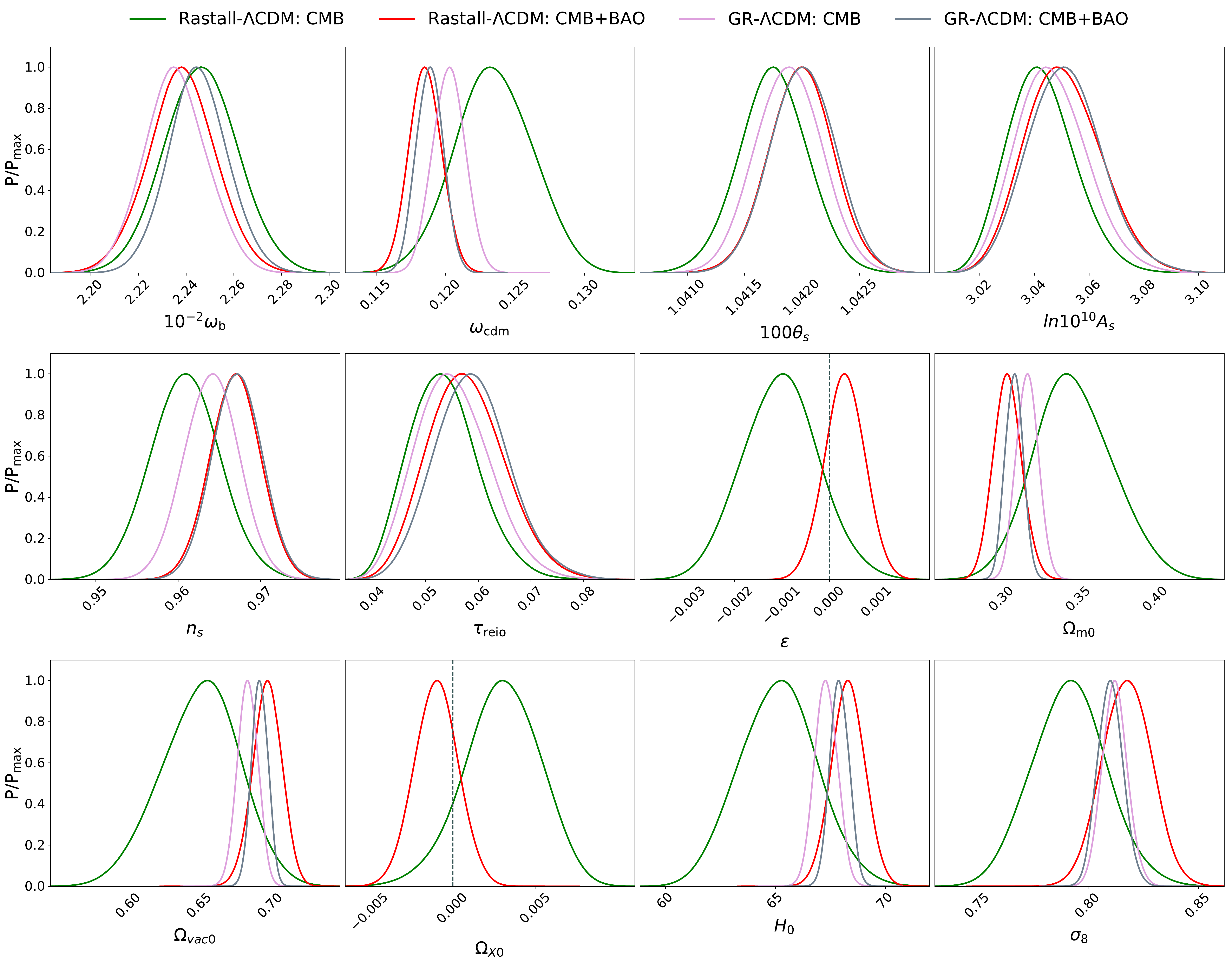}
\caption{One-dimensional marginalized distributions of the free and some derived parameters of the Rastall-$\Lambda$CDM and GR-$\Lambda$CDM models. }
\label{1D}
\end{figure*}

We have implemented the model in publicly available \texttt{CLASS} \cite{class} code, and used the \texttt{Multinest} \cite{feroz} algorithm in the parameter inference \texttt{Monte Python} \cite{monte} code with uniform priors on the model parameters to obtain correlated  Monte Carlo Markov Chain samples and Bayesian evidence. Further, we have used the \texttt{GetDist} Python package to analyze the samples. We obtain the observational constraints on all the Rastall-$\Lambda$CDM model parameters by using first only the CMB data and then the combined CMB+BAO data. For comparison purposes, we also show the constraints on the GR-$\Lambda$CDM model parameters. The CMB data set alone is known to well constrain the six baseline parameters of the GR-$\Lambda$CDM model. The Rastall-$\Lambda$CDM model carries an additional parameter, namely $\epsilon$. Therefore, we also combine the BAO data with CMB in order to obtain possibly tighter constraints on the Rastall-$\Lambda$CDM model parameters, and also to break any possible degeneracy of the new parameter $\epsilon$ with the other baseline parameters.

 \begin{figure*}[hbtt!]\centering
\includegraphics[width=14.5cm]{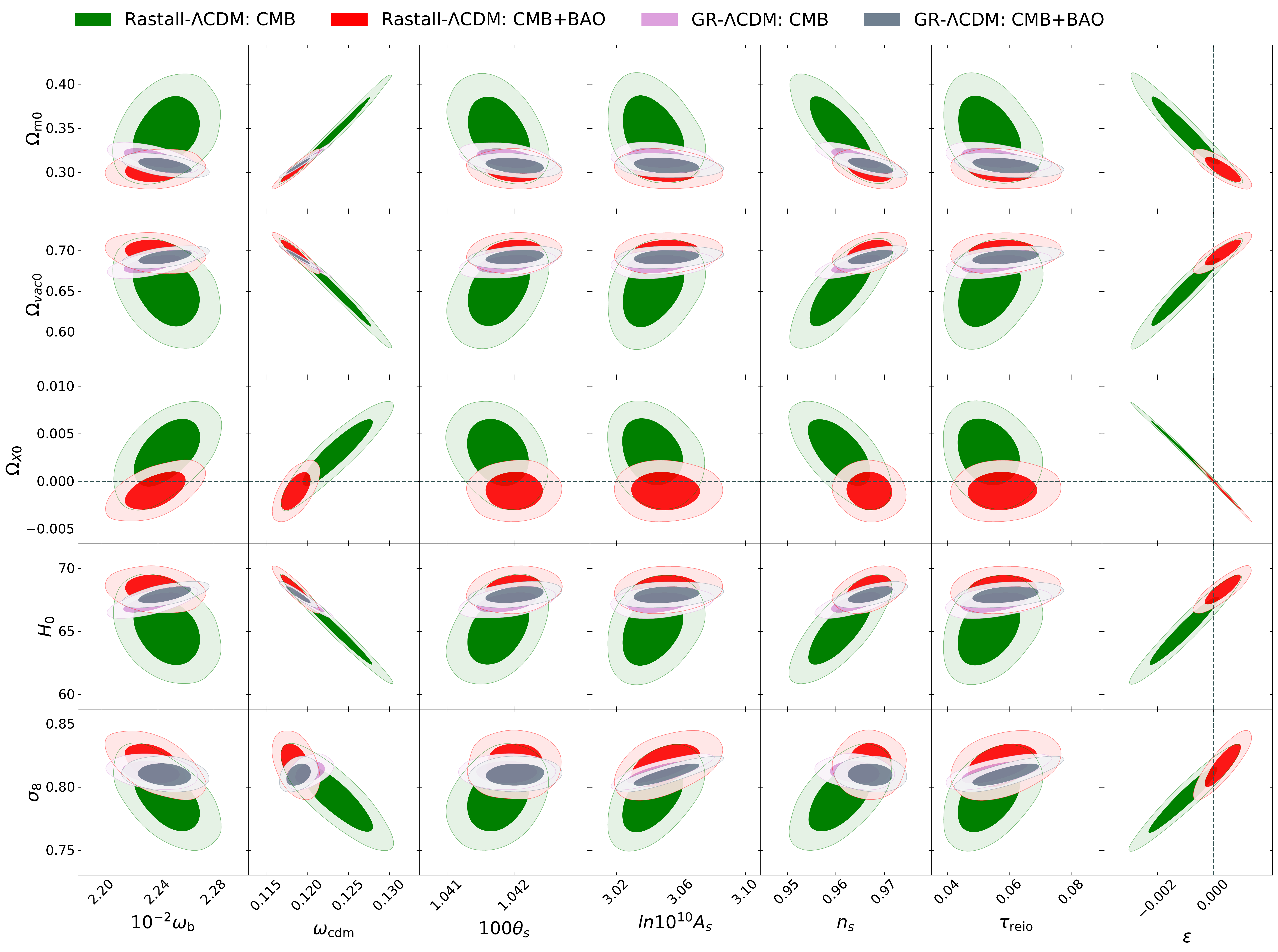} 
\caption{Two-dimensional (68\% and 95\% CLs) marginalized distributions  of the free and some derived parameters of the Rastall-$\Lambda$CDM  and GR-$\Lambda$CDM.} 
\label{2D}
 \end{figure*}

\begin{figure*}[hbtt!]\centering
\includegraphics[width=14.5cm]{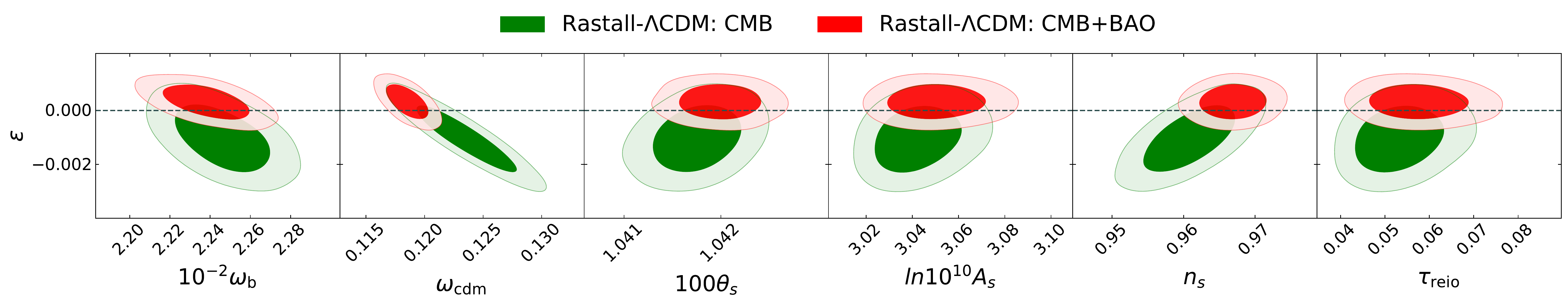} 
\caption{Two-dimensional (68\% and 95\% CLs) marginalized distributions of $\epsilon$ with the other baseline parameters of the Rastall-$\Lambda$CDM model.} 
\label{2D_eps}
 \end{figure*}
 
 Table \ref{tableI} displays the constraints, at 68\% and 95\% confidence levels (CLs), on the baseline seven free parameters and some derived parameters of the Rastall-$\Lambda$CDM model and, for comparison, on those of the GR-$\Lambda$CDM model, both from the CMB and combined CMB+BAO data sets. For these constraints presented for both models in the same table, Figure \ref{1D} shows the one-dimensional marginalized distributions and Figure \ref{2D} shows the two-dimensional (68\% and 95\% CLs) marginalized distributions of the derived parameters with regard to the baseline free parameters. From all these, we immediately notice that, as it is the case for the GR-$\Lambda$CDM model as well, the CMB+BAO data set puts tight constraints on the parameters of the Rastall-$\Lambda$CDM model, when compared to the constraints put by CMB data set alone. On the other hand, in contrast to the GR-$\Lambda$CDM model, when the BAO data set is not included, we notice larger error bounds (loose constraints) on some of the Rastall-$\Lambda$CDM model parameters, for instance, the error bounds of $\omega_{\rm cdm}$ and the derived parameters are larger in the Rastall-$\Lambda$CDM model compared to the GR-$\Lambda$CDM model, as a consequence of the deviation of the Rastall parameter $\epsilon$ from zero. It turns out that the parameter of our main concern, the Rastall parameter $\epsilon$ measuring the deviation from GR, is constrained as
\begin{equation*}
\begin{aligned}
    \epsilon&=-0.0010\pm 0.0008\pm 0.0015 \quad \textnormal{from CMB,}\\
    \epsilon&=0.0003\pm 0.0004\pm 0.0008 \quad \textnormal{from CMB+BAO},
    \end{aligned}
\end{equation*}
at 68\% and 95\% CLs. The constraints from the CMB data as well as the combined CMB+BAO data set suggest that, in line with our conclusion reached via a preliminary investigation in Sect.~\ref{sec:preliminary}, the Rastall parameter $\epsilon$ is well consistent with zero at 95\% CL, which in turn implies that there is no significant statistical evidence for deviation from GR via Rastall gravity. We note however that, as may be seen from the probability regions and mean values of $\epsilon$, the Rastall parameter prefers negative values in the case of CMB data while positive values when BAO data set is included (CMB+BAO data). Further, although they are mostly minor, the Rastall gravity extension of the standard $\Lambda$CDM has some consequences on the cosmological parameters, which deserve further discussion that could be informative about the features of the Rastall gravity and/or whether it is a promising modified gravity theory or not.

\begin{figure}[t!]\centering
\includegraphics[width=0.4\textwidth]{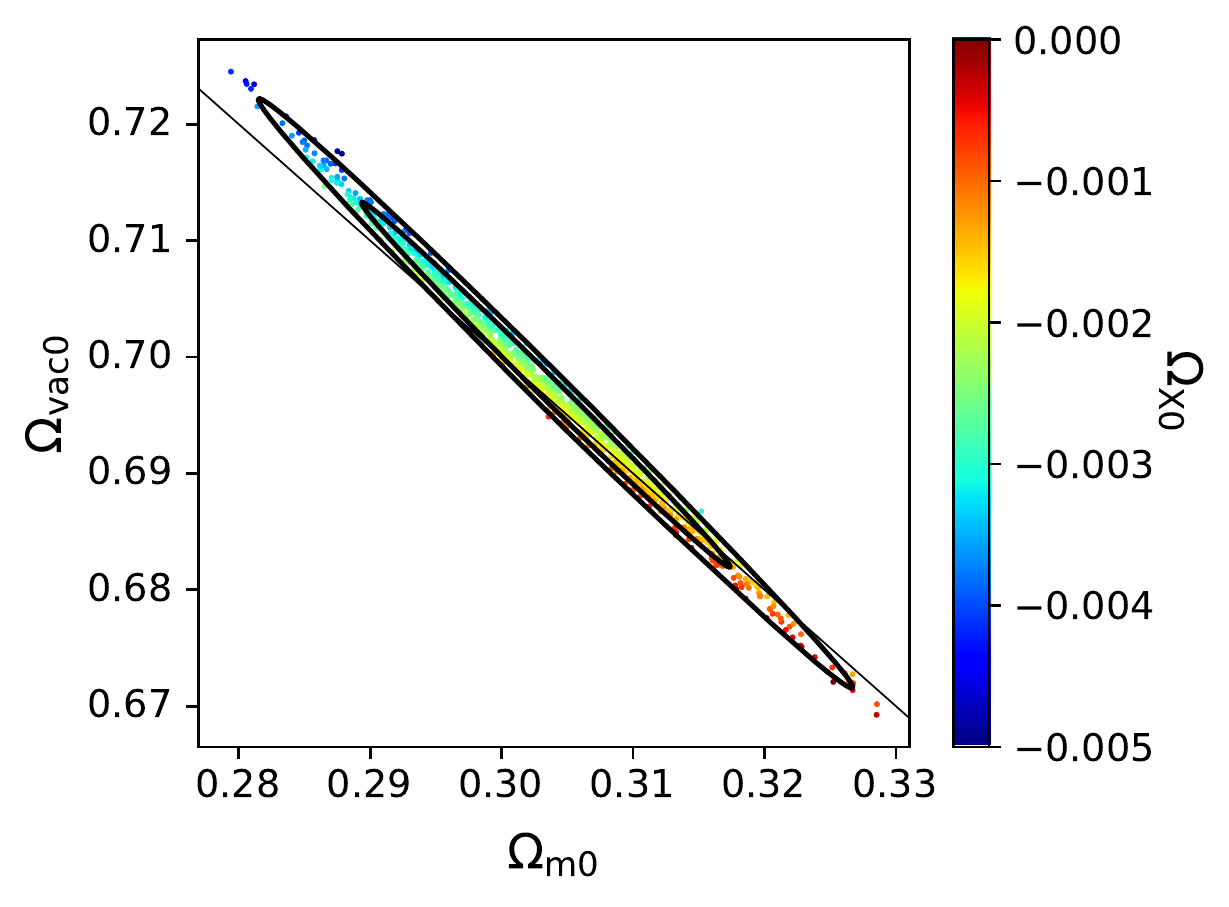} 
\caption{Two-dimensional posterior distributions of $\{\Omega_{\rm m0}, \Omega_{\rm vac0}\}$ colour coded by $\Omega_{\rm X0}$ with CMB+BAO data. The line across the contours is given by $\Omega_{\rm m0}+\Omega_{\rm vac0}=1$. } 
\label{fig:7}
 \end{figure}

In Figure \ref{2D_eps}, we present the two-dimensional (68\% and 95\% CLs) marginalized distributions that show how the six of the baseline parameters of the GR-$\Lambda$CDM model are affected by the Rastall gravity extension, i.e., $\epsilon$. We observe that $\epsilon$ is negatively correlated with $\omega_{\rm cdm}$ and $\omega_{\rm b}$ for both the CMB data and the combined CMB+BAO data sets. Other four parameters, $\theta_s$, $A_s$, $n_s$ and $\tau_{\rm reio }$ seem to have minor positive correlations with $\epsilon$ in the case of CMB data but no correlation in the case of CMB+BAO data. Accordingly, in Table \ref{tableI} and Figure \ref{1D}, one may notice the shifts in the mean values and one dimensional probability distributions of different parameters. Also,  see Figure \ref{2D} for the consequences of the Rastall gravity extension, i.e., $\epsilon$, via the two-dimensional marginalized distributions of the derived parameters with regard to the baseline free parameters. The last column of the same figure is of particular interest as it displays the two-dimensional marginalized distributions of the derived parameters with regard to the constraints on the Rastall parameter $\epsilon$. We notice that smaller values of $\epsilon$ lead to larger values of the present-day density parameter of matter $\Omega_{\rm m0}$ because of the negative correlation between these two parameters. It is in line with \eqref{eq:rhoi} which suggests that matter energy density dilutes less efficiently with time in a universe with negative values of $\epsilon$, and thereby leading to larger matter density parameter in the present-day Universe. Accordingly, in the case of CMB data, where $\epsilon$ has higher probability to lie in the negative range, we see higher values of $\Omega_{\rm m0}$ in the Rastall-$\Lambda$CDM when compared to the GR-$\Lambda$CDM. Namely, the CMB data set predicts $\Omega_{\rm m0}=0.347^{+ 0.024}_{-0.027}$ for the Rastall-$\Lambda$CDM model, while $\Omega_{\rm m0}=0.316^{+0.007}_{-0.007}$ for the GR-$\Lambda$CDM. On the other hand, the combined CMB+BAO data set prefers larger probability region of $\epsilon$ in the positive range, and thereby it predicts smaller values of $\Omega_{\rm m0}$ in the Rastall-$\Lambda$CDM when compared to the GR-$\Lambda$CDM. For, the combined CMB+BAO data set predicts $\Omega_{\rm m0}=0.304^{+ 0.009}_{-0.009}$ for the Rastall-$\Lambda$CDM model, while $\Omega_{\rm m0}=0.308^{+0.006}_{-0.006}$ for the GR-$\Lambda$CDM model, see Table \ref{tableI}. We notice a positive correlation between the present-day density parameter of the usual vacuum energy and the Rastall parameter for the both data sets. The CMB (CMB+BAO) data set favors smaller (larger) values of the density parameter of the usual vacuum energy density, viz., $\Omega_{\rm vac0}=0.650^{+0.029}_{-0.026}$ ($\Omega_{\rm vac0}=0.697^{+0.010}_{-0.010}$) for the Rastall-$\Lambda$CDM while $\Omega_{\rm vac0}=0.684^{+0.007}_{-0.007}$ ($\Omega_{\rm vac0}=0.692^{+0.006}_{-0.006}$)
for the GR-$\Lambda$CDM. This reduced (enhanced) amount of the usual vacuum energy density parameter is however compensated just slightly by that of the effective source (which behaves like a cosmological constant at $z\sim0$) as the CMB (CMB+BAO) data set favors its positive (negative) values owing to the almost perfect negative correlation between its present-day density parameter $\Omega_{\rm X0}$ and the Rastall parameter $\epsilon$. Namely, the constraint on total present-day density parameter of those of the usual vacuum energy and the effective source reads $\Omega_{\rm vac0}+\Omega_{\rm X0}=0.653^{+0.027}_{-0.024}$ (68\% CL) from the CMB data set, and $\Omega_{\rm vac0}+\Omega_{\rm X0}=0.696\pm 0.009$ (68\% CL) from the combined CMB+BAO data set. Figure \ref{fig:7} displays the two-dimensional posterior distributions of $\{\Omega_{\rm m0},\Omega_{\rm vac0}\}$ colour coded by the $\Omega_{\rm X0}$, at 68\% and 95\% CLs for the combined CMB+BAO data set. Here, we notice that the posterior distribution sample points of $\Omega_{\rm X0}$ (and the corresponding contours for 68\% and 95\% CLs) pretty much cluster around a line that deviates from the GR-$\Lambda$CDM line $\Omega_{\rm m0}+\Omega_{\rm vac0}=1$, due to the presence of $\Omega_{\rm X0}$.

\begin{figure}[t!]\centering
\includegraphics[width=0.45\textwidth]{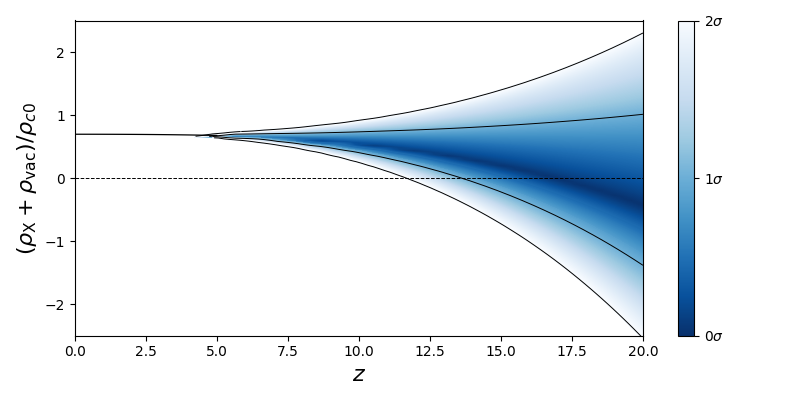} 
\caption{Plot of $(\rho_{\rm X}+\rho_{\rm vac})/\rho_{\rm c0}$ vs $z$ with 68\% and 95\% error regions in the case of CMB+BAO data.} 
\label{fig:zstar}
 \end{figure}

\begin{figure}[t!]\centering
\includegraphics[width=0.38\textwidth]{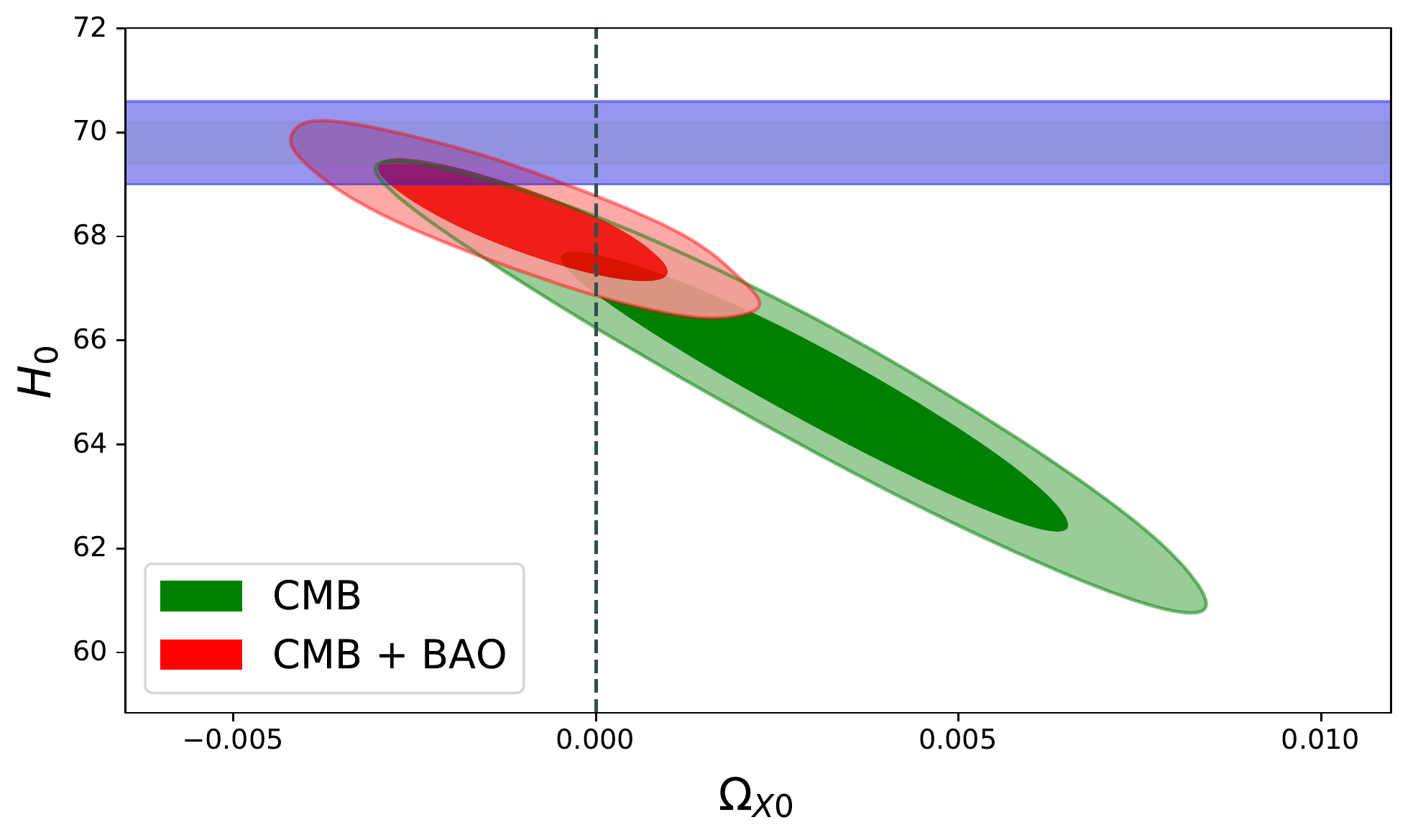} 
\caption{Two-dimensional (68\% and 95\% CLs) marginalized distributions of $H_0$ versus $\Omega_{\rm X0}$ for the Rastall-$\Lambda$CDM model. The horizontal blue band is for the model independent TRGB $H_0$ measurement $H_0= 69.8 \pm 0.8\, {\rm km\,s}^{-1}{\rm Mpc}^{-1}$ \cite{Freedman:2019jwv}.} 
\label{fig:H0}
 \end{figure}

\begin{figure}[t!]\centering
\includegraphics[width=0.46\textwidth]{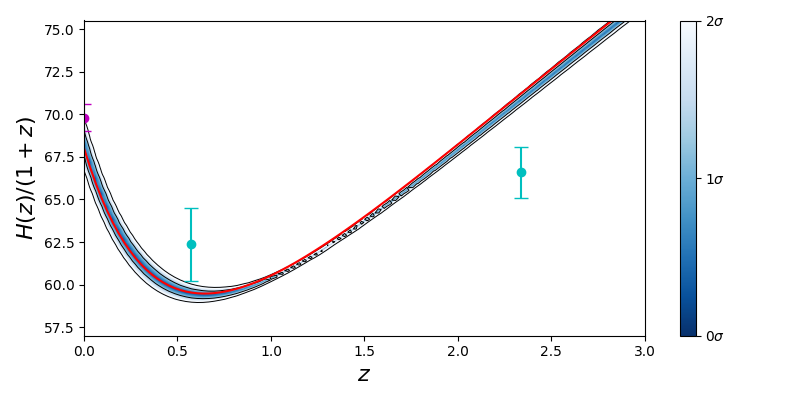} 
\caption{$H(z)/(1+z)$ vs $z$ with 68\% and 95\% error regions in the case of CMB+BAO data. Here, the red curve stands for the GR-$\Lambda$CDM model corresponding to the mean values of the parameters. The three error bars stand for  $H_0= 69.8 \pm 0.8\, {\rm km\,s}^{-1}{\rm Mpc}^{-1}$ from the TRGB $H_0$ \cite{Freedman:2019jwv}, $H(z=0.57) = 97.9 \pm 3.4\,{\rm km\,s}^{-1}{\rm Mpc}^{-1}$ \cite{bao3}, and $H(z=2.34) = 222.4 \pm 5.0\,{\rm km\,s}^{-1}{\rm Mpc}^{-1}$ from the latest BAO data \cite{Delubac:2014aqe}.} 
\label{fig:8a}
 \end{figure}

We note that while the CMB data set by alone favors positive energy density values for the effective source $\rho_{\rm X}$ accompanying to the actual energy sources due to the Rastall gravity extension, the CMB+BAO data set including BAO data (relatively low-redshift data compared to the CMB data set) favors negative energy density values for it, namely, $\Omega_{\rm X0}=0.0030\pm0.0023$ (68\% CL) from the CMB data set, and $\Omega_{\rm X0}=-0.0010\pm0.0013$ (68\% CL) from the combined CMB+BAO data set. This shows that, when the BAO data set is included, the effective source $\rho_{\rm X}$ indeed screens the usual vacuum energy at finite redshift and that the Rastall gravity may be counted among the cosmological models \cite{Delubac:2014aqe,Aubourg:2014yra,Sahni:2014ooa,Mortsell:2018mfj,Poulin:2018zxs,Capozziello:2018jya,Wang:2018fng,Dutta:2018vmq,Banihashemi:2018oxo,Banihashemi:2018has,Visinelli:2019qqu,Akarsu:2019hmw,Ye:2020btb,Perez:2020cwa} that were suggested for alleviating a number of persistent low-redshift tensions, including the $H_0$ tension, by a dynamical DE that assumes negative energy density values (or by a mechanism dynamically screening the cosmological constant) at finite redshift. In Figure \ref{fig:zstar}, in order to visualize the screening mechanism, we show the evolution of the total energy density  of the  effective source plus the usual vacuum energy (scaled to the present-day critical energy density), $(\rho_{\rm X}+\rho_{\rm vac})/\rho_{\rm c0}$, versus the redshift with probability regions up to 95\% CL (the darker the more probable), using the \texttt{fgivenx} python package \cite{handley19}. We see that the effective source completely screens the usual vacuum energy, $\rho_{\rm X}+\rho_{\rm vac}=0$, at a redshift $z_*> 13.65$  (68\% CL) and $z_*> 11.68$ (95\% CL).
 However, these are too large compared to the $z_*$ values suggested in, e.g., Refs. \cite{Aubourg:2014yra,Sahni:2014ooa,Akarsu:2019hmw}. This might be signaling that this mechanism does not work efficiently enough in the Rastall-$\Lambda$CDM model. For instance, we indeed observe almost a perfect positive correlation between $H_0$ and $\epsilon$, which implies that larger values of $\epsilon$ would correspond to larger values of $H_0$. We see in Table \ref{tableI} that, in comparison to the GR-$\Lambda$CDM model, the combined CMB+BAO data set favors a slightly larger mean value for $H_0$ in the Rastall-$\Lambda$CDM model, which seems to be an improvement for a better agreement with, e.g., the model independent $H_0$ values measured from the distance ladder measurements (e.g., $H_0= 69.8\pm 0.8$ from a recent calibration of the TRGB applied to Type Ia supernovae \cite{Freedman:2019jwv}), see Figures \ref{fig:H0} and \ref{fig:8a}.
 However, a more careful look reveals that this improvement is not robust. The combined CMB+BAO data set predicts $H_0=68.31^{+ 0.76+1.50}_{-0.76-1.50}\,{\rm km\,s}^{-1}{\rm Mpc}^{-1}$ for the Rastall-$\Lambda$CDM model, while $H_0=67.92^{+0.43+0.83}_{-0.43-0.82}\,{\rm km\,s}^{-1}{\rm Mpc}^{-1}$ for the GR-$\Lambda$CDM model. We note that, in contrast to the GR-$\Lambda$CDM model, $H_0$ in the case of the Rastall-$\Lambda$CDM model, even at $68\%$ CL, agrees with the TRGB $H_0$ value. This may be found promising, but, this is because of the large widening in the one-dimensional marginalized probability distribution of $H_0$, viz., largely increased errors, while a minor shift to the larger mean value of $H_0$. This shows that the Rastall-$\Lambda$CDM model offers a relaxation to the so called $H_0$ tension prevailing within the GR-$\Lambda$CDM model, though not robustly. In figure \ref{fig:8a}, we plot $H(z)/(1+z)$ to display the situation with respect to TRGB $H_0$ and BAO data by considering the constraints from the combined CMB+BAO data. We see only a slightly better representation of the three data points by the Rastall-$\Lambda$CDM model.

 It seems that, as we have discussed in Sect.~\ref{sec:preliminary}, the screening mechanism provided via the effective source due to the Rastall gravity when $\epsilon>0$ does not work efficiently as the altered redshift dependence of dust (due to the violation of the local energy-momentum conservation) opposes that by keeping $\epsilon$ close to zero, and the CMB data set (as well as the Ly-$\alpha$ data, see the discussion in Sect.~\ref{sec:preliminary}) tends $\epsilon$ towards negative values. Next, we observe that both data sets lead to a positive correlation of $\epsilon$ with $\sigma_8$, similar to $H_0$. Therefore, the larger (smaller) values of $\epsilon$ would correspond to larger (smaller) values of both $H_0$ and $\sigma_8$, which in turn implies that both the so called $H_0$- and $\sigma_8$-tensions prevailing within the GR-$\Lambda$CDM model cannot be relaxed together in the Rastall-$\Lambda$CDM model (see, e.g., \cite{Kumar:2019wfs} achieving a simultaneous relaxation of the two tensions by a non-minimal interaction in the dark sector).
 
\begin{figure}[h!]\centering
\includegraphics[width=0.35\textwidth]{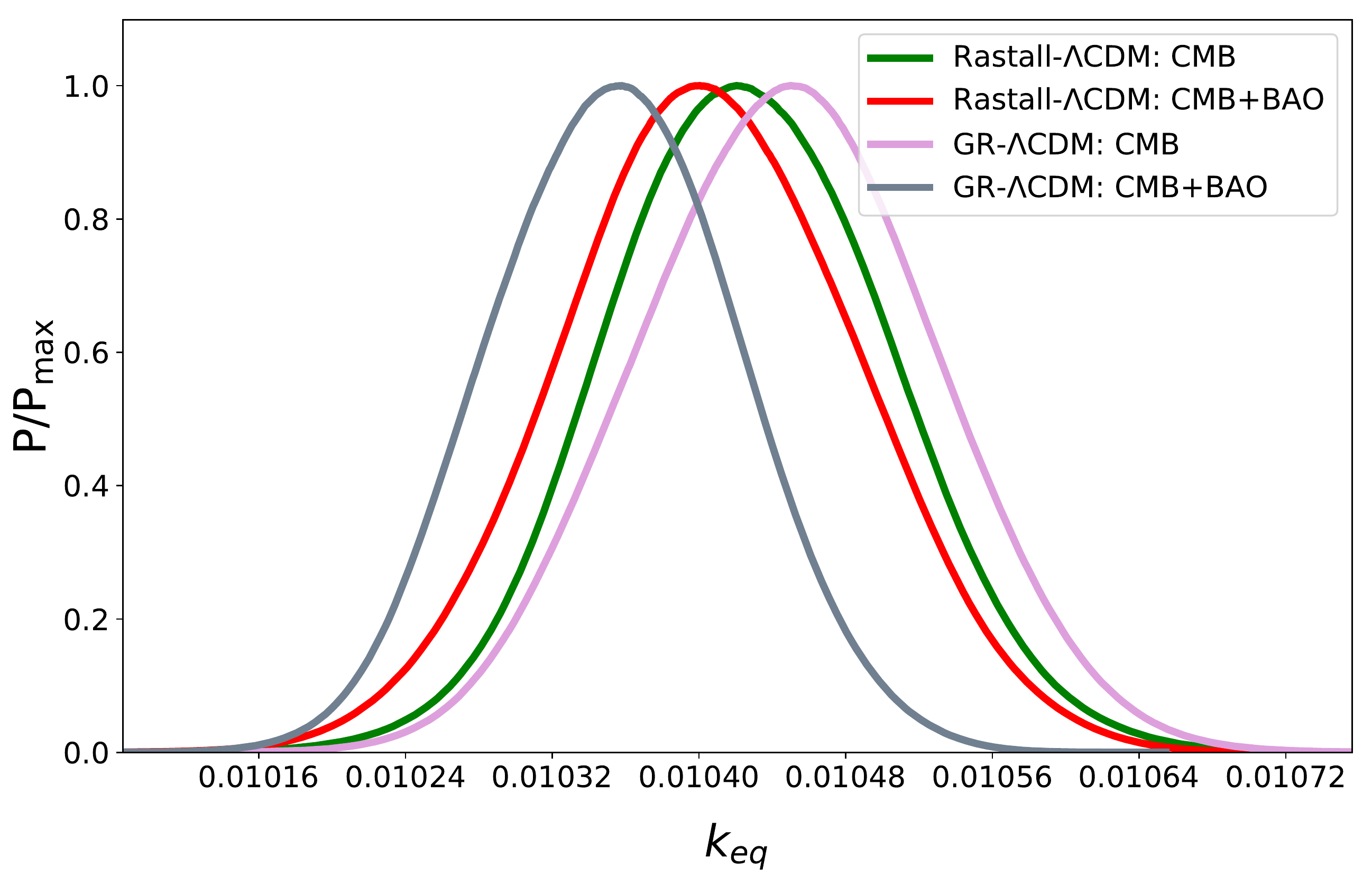} 
\caption{One-dimensional marginalized distribution of $k_{\rm eq}$ for the Rastall-$\Lambda$CDM and GR-$\Lambda$CDM models.} 
\label{fig:k_eq}
 \end{figure}

Finally, we have a look at the situation of the wavenumber of a mode that enters the horizon at the radiation-matter transition, $k_{\rm eq}$, which is a highly sensitive parameter to the modifications to GR, and related to the dynamics of the Universe at a redshift larger than the redshifts related to the CMB data. In the case of the Rastall-$\Lambda$CDM model, we find the constraints $ k_{\rm eq}=0.010427\pm 0.000078\,{\rm Mpc}^{-1}$ from CMB data, and $k_{\rm eq}=0.010406\pm 0.000081\,{\rm Mpc}^{-1}$ from CMB+BAO data. On the other hand, for the GR-$\Lambda$CDM model, CMB data provide $k_{\rm eq}=0.010448\pm 0.000081\,{\rm Mpc}^{-1}$, and the CMB+BAO data provide  $k_{\rm eq}=0.010354\pm 0.000068\,{\rm Mpc}^{-1}$. Figure \ref{fig:k_eq} shows one dimensional marginalized distributions of $k_{\rm eq}$ in the four cases. We notice that $k_{\rm eq}$ probability curves of the Rastall-$\Lambda$CDM model are bit together but we see a larger shift among the ones for the GR-$\Lambda$CDM model when compared to the Rastall-$\Lambda$CDM model. Though this shift is not significant, but may be signaling a better consistency of the Rastall-$\Lambda$CDM model with both the data sets in comparison to the GR-$\Lambda$CDM model.
 
 For a better understanding of the results we obtained for the Rastall-$\Lambda$CDM model, it may be useful to elucidate the reason behind the fact that the CMB data (CMB+BAO data) favor smaller (larger) $H_0$ values compared to the GR-$\Lambda$CDM model. As mentioned in the introduction, the constraint on $H_0$ from CMB data depends on the angular scale on the sky, $\theta_{\star} = r_{\star}/D_{\rm M}$, where $r_{\star}=\int^{\infty}_{z_{\star}}c_sH^{-1}{\rm d}z$ is the comoving sound horizon at recombination---determined by the pre-recombination ($z>z_{\star}$) physics---and $D_{\rm M}=\int_0^{z_{\star}}H^{-1}{\rm d}z$ is the comoving angular diameter distance at recombination---determined by the post-recombination ($z<z_{\star}$) evolution of $H(z)$ \cite{Dodelson03}. We notice that none of the terms in \eqref{eq:fried_3} scale faster than the radiation (within the allowed region for $\epsilon$ by the observations, see Table \ref{tableI} for the constraints) and $k_{\rm eq}$ is almost the same in both models---whether the constraints from the CMB data or the combined CMB+BAO are considered---, which in turn imply almost the same $r_{\star}$ value in all cases as the pre-combination physics/dynamics does not differ in any case---recall that the radiation dominated Universe solutions do not differ in the Rastall gravity and GR. Also, Planck satellite measures $\theta_{\star}$ (cf. $\theta_{\rm s}$, see Table \ref{tableI} for constraints) very robustly and almost independently of the cosmological model with a high precision. Therefore, $D_{\rm M}$ must be almost the same in all cases. Now, we notice that the CMB data by alone favor slightly negative values of $\epsilon$ accompanied by an enhancement in $\omega_{\rm cdm}$, which then result in an enhancement in $\Omega_{\rm m0}$ and positive $\Omega_{\rm X0}$, see, e.g. Figure \ref{1D}. One may check that,  in comparison to the GR-$\Lambda$CDM when the constraints on the parameters given in Table \ref{tableI} are considered, despite the fact that the slightly negative mean value of $\epsilon$ leads the dust-like contributions in $H(z)$ [see \eqref{eq:fried_3}] to scale slightly slower than the usual $(1+z)^3$ dependence, the enhanced $\Omega_{\rm m0}$ value and the positivity of $\Omega_{\rm X0}$---i.e., reduced $\Omega_{\rm vac0}$---lead $H(z)$ to grow faster for $z<z_{\star}$ and become larger for the interval $0\lesssim z<z_{\star}$. The compensation of this enhanced $H(z)$ in the interval $0\lesssim z<z_{\star}$ in turn implies a diminished $H(z)$ for $z\sim 0$---a smaller $H_0$ as well---to keep $D_{\rm M}$ (almost) unaltered. This situation leads to a strong correlation between $H_0$ and $\epsilon$, which can be seen in Figure \ref{2D}. The inclusion of the BAO data, in addition to the CMB data, in our analyses not only decreases the errors on $\epsilon$ but also brings its mean value to a value just slightly larger than zero, which in turn  yields smaller errors on $H_0$ and its slightly larger mean value than the one predicted within the GR-$\Lambda$CDM model, though it is not significant enough to relax the so-called $H_0$ problem (see Table \ref{tableI}, and Figures \ref{1D} and \ref{fig:H0}).

In the case of CMB data, we notice  larger error regions of all the six baseline parameters (except $\theta_s$, $A_s$ and $\tau_{\rm reio }$) and other derived parameters in the Rastall-$\Lambda$CDM model when compared to the GR-$\Lambda$CDM model. Indeed, the additional parameter $\epsilon$ in the Rastall-$\Lambda$CDM model penalizes the statistical fit of this model to the data when compared to the GR-$\Lambda$CDM model. However, inclusion of BAO data, that is, CMB+BAO data put tight constraints on all the model parameters, and thereby reduce the error regions of the parameters considerably in both the models. In the following section, we calculate Bayesian evidences of the two models, and thereby do a comparison of the statistical fit.

\section{Bayesian evidence of the fit}
\label{sec:bayes}
For comparing statistical fit of the models under consideration in this work to the observational data, we use Bayesian evidence. In this regard Bayes' theorem reads 
 \begin{equation}\label{bayes}
		P(\Theta|D, \mathcal{M}) = \frac{\mathcal{L}(D|\Theta, \mathcal{M}) \pi(\Theta|\mathcal{M})}{\mathcal{E}(D|\mathcal{M})},
\end{equation}
for a given model $\mathcal{M}$ with the set of parameters $\Theta$ and the cosmological data $D$. Here, $P(\Theta|D, \mathcal{M})$ is the posterior distribution of the parameters $\Theta$; $\mathcal{L}(D|\Theta, \mathcal{M})$ is the likelihood function; $\pi(\Theta|\mathcal{M})$ is the prior probability of the model parameters, and $\mathcal{E}(D|\mathcal{M})$ is the Bayesian evidence calculated as
\begin{equation}\label{evidence}
\mathcal{E}(D|\mathcal{M}) = \int_\mathcal{M} \mathcal{L}(D|\Theta, \mathcal{M}) \pi(\Theta|\mathcal{M}) \text{d}\Theta.
\end{equation}

Further, we compute the ratio of the posterior probabilities for a model $\mathcal{M}_a$ with respect to a reference model $\mathcal{M}_b$ as
\begin{equation}
\frac{P(\mathcal{M}_a|D)}{P(\mathcal{M}_b|D)} = B_{ab}\frac{P(\mathcal{M}_a)}{P(\mathcal{M}_b)},
\end{equation}
where $B_{ab}$ is the Bayes' factor, evaluated as
\begin{equation}\label{bayes_factor}
B_{ab} = \frac{\mathcal{E}_a}{\mathcal{E}_b}.
\end{equation}

The Jeffreys' scale \cite{Jeffreys} is used to interpret the Bayes' factor by calculating $|\ln B_{ab}|$. The value of $|\ln B_{ab}|$ lying in the range [0,1) implies the strength of the evidence to be weak or inconclusive, while a definite or positive evidence is implied by the values in the range [1,3). Further, the strength of the evidence is strong for $|\ln B_{ab}|$ lying in [3,5), and is the strongest for $|\ln B_{ab}|$ greater than 5. 
Table \ref{tableII} displays the Bayesian evidence of Rastall-$\Lambda$CDM in comparison with the GR-$\Lambda$CDM model in the case of CMB and CMB+BAO data, where $\mathcal{E}_{\rm Rastall}$ and $\mathcal{E}_{\rm GR}$, respectively stand for the Bayesian evidences of the Rastall-$\Lambda$CDM and GR-$\Lambda$CDM models. We observe a definite evidence ($|\ln \mathcal{E}_{\rm Rastall, \rm GR}|\in [1,3)$) in the case of CMB data, whereas weak evidence ($|\ln \mathcal{E}_{\rm Rastall, \rm GR}|\in [0,1)$) is observed in the case of the  CMB+BAO data. 
Thus, the GR-$\Lambda$CDM model finds a better fit to the CMB data in comparison to the Rastall-$\Lambda$CDM model, but the weak evidence in the case of the combined CMB+BAO data, suggests that both the models fit equally well to the CMB+BAO data, as expected.

\begin{table}[h!]
\small
\caption{Bayesian evidences of the Rastall-$\Lambda$CDM and GR-$\Lambda$CDM models, where $\ln \mathcal{E}_{\rm Rastall, \rm GR}=\ln \mathcal{E}_{\rm Rastall}-\ln\mathcal{E}_{\rm GR}$.}
\label{tableII}
\setlength\extrarowheight{2pt}
\begin{tabular} { |l| l| l| l| l|  }  \hline 
   &  CMB     & CMB+BAO        \\ 
\hline
$\ln \mathcal{E}_{\rm Rastall} $ & $-1407.02\pm 0.21$  & $-1413.67\pm 0.21$  \\[1ex] 
  \hline
 $\ln \mathcal{E}_{\rm GR}$ &  $-1404.56\pm 0.19$ & $-1413.30\pm 0.19$  \\[1ex] 
    
  \hline
  
   $\ln \mathcal{E}_{\rm Rastall, \rm GR}$ &  $-2.46\pm 0.28$ & $-0.37\pm 0.28$  \\[1ex] 
    
  \hline 
 
\end{tabular}
\end{table}

\section{Conclusions}
\label{sec:Conclusions}

We have constructed the extension of the standard $\Lambda$CDM model (GR-$\Lambda$CDM) by switching the gravity theory from GR to the Rastall gravity (Rastall-$\Lambda$CDM), see Sect.~\ref{sec:cosmo} . We then have reviewed it---via a preliminary investigation of its features for a demonstration of how it works, and a guide to the values of its parameters---in a proper manner, namely, in a manner clearly identifying and handling the two simultaneous modifications of different nature in the material content side of the Einstein field equations of GR, see Sects.~\ref{sec:cosmo}, \ref {sec:GRequiv} and \ref{sec:preliminary}. It has then turned out that it is not possible to reach a decisive conclusion only through a preliminary investigation, for instance, whether the $H_0$ and/or BAO data show tendency of $\epsilon$ deviating from zero (GR) in a certain direction. Further, we also have learned that a significant improvement with regard to $H_0$ and/or BAO data would most likely lead to spoiling of a successful description of the early Universe, which signals that CMB data would keep $\epsilon$ values close to zero. These inspections have led us to expect only an insignificant deviation from the standard $\Lambda$CDM model when it is extended from GR via the Rastall gravity, and persuaded us that a conclusive answer cannot be given unless the model is rigorously confronted/constrained with the observational data.

Considering the background and perturbation dynamics (Sect.~\ref{sec:pert}), we have explored the full parameter space of the Rastall-$\Lambda$CDM model---viz., the Rastall gravity extension of the six-parameter base GR-$\Lambda$CDM  described by the additional parameter $\epsilon$---using the latest CMB data set as well as the latest combined CMB+BAO data set, see Sect.~\ref{sec:obsanalysis}.  Also, for comparison, we have presented the corresponding constraints/results on the GR-$\Lambda$CDM model. It turned out that, as it is the case for the GR-$\Lambda$CDM model as well, the CMB+BAO data set puts tight constraints on the parameters of the Rastall-$\Lambda$CDM model and that, in contrast to the case for the GR-$\Lambda$CDM model, the CMB data set by alone puts loose constraints (larger error bounds) on some of the Rastall-$\Lambda$CDM model parameters---particularly, on the dimensionless density of cold dark matter $\omega_{\rm cdm}$ and all of the derived parameters---as a consequence of a wider range of deviation of the Rastall parameter $\epsilon$ from zero. Yet, both analyses suggest that, in line with our conclusion reached via a preliminary investigation in Sect.~\ref{sec:preliminary}, the Rastall parameter $\epsilon$ is well consistent with zero at 95\% CL, which in turn implies that there is no significant statistical evidence for deviation from GR via the Rastall gravity. We note however that, as may be seen from the probability regions and mean values of $\epsilon$, the Rastall parameter prefers negative values in the case of CMB data while positive values when BAO data set is included (CMB+BAO data). Despite the fact that they are basically minor within the allowed small range of the Rastall parameter from the data, we have explored the consequences of/tendencies led by the Rastall gravity on the cosmological parameters in the light of the observational analyses. Our results can be a guide for the research community that studies the Rastall gravity in various aspects of gravitation and cosmology, where, in general, as we have found in this work that the Rastall parameter cannot be out of the range $-0.0001 < \epsilon < 0.0007$ at 68\% CL. Being that range an observational boundary imposed from high precision full CMB data set along with the BAO data set, it, in principle, must be obeyed as a new bound in any qualitative study within this modified theory of gravity. Finally---in support of our conclusions here---comparing statistical fit of these two models to the observational data by using Bayesian evidence, the GR-$\Lambda$CDM model finds a better fit to the CMB data in comparison to the Rastall-$\Lambda$CDM model, but the weak evidence in the case of the combined CMB+BAO data, suggests that both the models fit equally well to the CMB+BAO data.

It also is worth mentioning as one of our conclusions that, if we assume that the standard physical ingredients of the Universe considered here are the true physical ingredients of the actual Universe, our finding that the term $g_{\mu\nu}R$ contributes to the Einstein field equations \eqref{eq:fieldeq} with a coefficient in the range $(-0.5001,-0.4993)$ from the combined CMB+BAO data at 68\% CL, i.e., a coefficient equal to $-1/2$ with a precision of $\mathcal{O}(10^{-4})$, can be taken as another new demonstration of the power of general relativity (which guarantees the local conservation of the total energy-momentum tensor relying on the twice-contracted Bianchi identity).

On the other hand, the Rastall gravity, in fact, possesses interesting features that could be of interest in the context of cosmology, for instance, to address some of the tensions prevailing within the standard $\Lambda$CDM model based on GR. One particular example may be that, for positive values of $\epsilon$, the effective source arising due to the Rastall gravity assumes negative energy density values and screens the usual vacuum energy in line with Refs. \cite{Delubac:2014aqe,Aubourg:2014yra,Sahni:2014ooa,Mortsell:2018mfj,Poulin:2018zxs,Capozziello:2018jya,Wang:2018fng,Dutta:2018vmq,Banihashemi:2018oxo,Banihashemi:2018has,Visinelli:2019qqu,Akarsu:2019hmw,Ye:2020btb,Perez:2020cwa}, which suggest such a scenario for alleviating a number of persistent low-redshift tensions, including the so called Hubble constant $H_0$ tension (deficiency). Indeed, our observational analyses show an almost perfect positive correlation between $H_0$ and $\epsilon$, which implies that larger values of $\epsilon$ would correspond to larger values of $H_0$. And, as the combined CMB+BAO data set favors slightly positive values of $\epsilon$, this feature of the model works in the right direction and leads to predictions of larger $H_0$ values (compared to the GR-$\Lambda$CDM model) consistent with, for instance, the model independent TRGB $H_0$ measurements. However, a more careful look revealed that this improvement is not robust as it arises from the large widening in the one-dimensional marginalized probability distribution of $H_0$, viz., largely increased errors, while a minor shift to the larger mean value of $H_0$. We remind that the effective source comes along with a modification in the energy density redshift dependence of the actual matter source (viz., CDM) due to the EMT non-conservation feature of the Rastall gravity. Such a modification would obviously be more and more effective on the dynamics of the Universe with the increasing redshift. Therefore, in the case of the combined CMB+BAO data, it is conceivable that the high redshift data (viz., the CMB data relevant to $z\sim1100$), in particular, tend to keep the redshift dependence of the actual matter source very close to its usual $(1+z)^3$ dependence, i.e, $\epsilon$ values almost equal to zero, and then does not allow the Rastall gravity to successfully realize a scenario wherein the usual vacuum energy is dynamically screened by the effective source. The lesson we learned from this is that, as we have seen the first signs in this direction in our preliminary investigations of the Rastall-$\Lambda$CDM model, the two simultaneous modifications of different nature in the material content side of the standard Einstein field equations, arising from the relaxation of the contribution of the term $g_{\mu\nu}R$ on the spacetime geometry side, act against each other. And, through a further modification of the Rastall gravity, it could probably be possible to reach a new modified theory of gravity which is relaxed from such a dichotomy between the two simultaneous modifications of different nature in the material content side of the standard Einstein field equations arising from a modification in the spacetime geometry side. We find this result important as such situations may exist in some other similar type of modified gravity theories.

\begin{acknowledgments}
We thank the referees for criticisms that helped us to improve the paper. \"{O}.A. acknowledges the support by the Turkish Academy of Sciences in scheme of the Outstanding Young Scientist Award  (T\"{U}BA-GEB\.{I}P). N.K. acknowledges the post-doctoral research support from the Istanbul Technical University (ITU). N.K. also acknowledges the COST Action CA15117 (CANTATA). S.K. and S.S. gratefully acknowledge the support from SERB-DST project No. EMR/2016/000258. R.C.N. would like to thank the Brazilian agency FAPESP for financial support under Project No. 2018/18036-5.
\end{acknowledgments}



\begin{thebibliography}{99}
\bibitem{footnote}{As we mentioned above, if baryons evolve following the modification rules imposed by Rastall gravity on GR, then the full Boltzmann equations for the baryons should also be modified. Also, universal couplings generally require screening mechanisms to protect baryonic interactions in high density environments, where the local gravity measurements are tightly constrained.  Also, non-standard dynamics of the ordinary particles is tightly constrained from the BBN prevision \cite{B1,B2,B3,B4}. However, the relevant investigation  of these various factors on physics of baryons is beyond the scope of the current paper and deserves to be investigated in detail in some future work. Thus, in the present paper, the baryons are assumed to be minimally coupled to gravity.}
\end{thebibliography}
\end{document}